\documentclass[11pt]{article}
\usepackage{amsmath}
\usepackage{graphicx}
\usepackage{comment}
\usepackage{natbib}
\usepackage{xcolor}
\usepackage{hyperref}
\usepackage{booktabs}
\usepackage{siunitx}

\usepackage{orcidlink}
\usepackage{amssymb}
\usepackage{authblk}
\usepackage{subcaption}
\usepackage[symbol]{footmisc}
\usepackage{float}

\usepackage[
  left=2cm,
  right=2cm,
  top=2.5cm,
  bottom=2.5cm
]{geometry}

\usepackage{lineno}
\title{KM3NeT/ARCA stacking search for high-energy neutrino point sources: the case of ultra-luminous infrared galaxies and blazars\\[0.5cm] 
\Large (The KM3NeT Collaboration\thanks{Corresponding Author: km3net-pc@km3net.de} )}

\author[b,a]{O.~Adriani\,\orcidlink{0000-0002-3592-0654}}
\author[c,bi]{A.~Albert}
\author[d]{A.\,R.~Alhebsi\,\orcidlink{0009-0002-7320-7638}}
\author[d]{S.~Alshalloudi\,\orcidlink{0009-0000-6757-7224}}
\author[f,e]{A. Ambrosone\thanks{Corresponding Author: antonio.ambrosone@unina.it}}
\author[g]{F.~Ameli}
\author[h]{F.~Andersen}
\author[i]{M.~Andre}
\author[j]{L.~Aphecetche\,\orcidlink{0000-0001-7662-3878}}
\author[k]{M. Ardid\,\orcidlink{0000-0002-3199-594X}}
% ...
\affil[a]{Universit{\`a} di Firenze, Dipartimento di Fisica e Astronomia, via Sansone 1, Sesto Fiorentino, 50019 Italy}
\affil[b]{INFN, Sezione di Firenze, via Sansone 1, Sesto Fiorentino, 50019 Italy}
\affil[c]{Universit{\'e}~de~Strasbourg,~CNRS,~IPHC~UMR~7178,~F-67000~Strasbourg,~France}
\affil[d]{Khalifa University of Science and Technology, Department of Physics, PO Box 127788, Abu Dhabi,   United Arab Emirates}
\affil[e]{INFN, Sezione di Napoli, Complesso Universitario di Monte S. Angelo, Via Cintia ed. G, Napoli, 80126 Italy}
\affil[f]{Universit{\`a} di Napoli ``Federico II'', Dip. Scienze Fisiche ``E. Pancini'', Complesso Universitario di Monte S. Angelo, Via Cintia ed. G, Napoli, 80126 Italy}
\affil[g]{INFN, Sezione di Roma, Piazzale Aldo Moro, 2 - c/o Dipartimento di Fisica, Edificio, G.Marconi, Roma, 00185 Italy}
\affil[h]{IFIC - Instituto de F{\'\i}sica Corpuscular (CSIC - Universitat de Val{\`e}ncia), c/Catedr{\'a}tico Jos{\'e} Beltr{\'a}n, 2, 46980 Paterna, Valencia, Spain}
\affil[i]{Universitat Polit{\`e}cnica de Catalunya, Laboratori d'Aplicacions Bioac{\'u}stiques, Centre Tecnol{\`o}gic de Vilanova i la Geltr{\'u}, Avda. Rambla Exposici{\'o}, s/n, Vilanova i la Geltr{\'u}, 08800 Spain}
\affil[j]{Subatech, IMT Atlantique, IN2P3-CNRS, Nantes Universit{\'e}, 4 rue Alfred Kastler - La Chantrerie, Nantes, BP 20722 44307 France}
\affil[k]{Universitat Polit{\`e}cnica de Val{\`e}ncia, Instituto de Investigaci{\'o}n para la Gesti{\'o}n Integrada de las Zonas Costeras, C/ Paranimf, 1, Gandia, 46730 Spain}
% ...
\affil[bi]{Universit{\'e} de Haute Alsace, rue des Fr{\`e}res Lumi{\`e}re, 68093 Mulhouse Cedex, France}
% ...

\date{}

\begin{document}

\maketitle

\begin{abstract}
A stacking  search for high-energy astrophysical neutrino emission is presented using three candidate source populations: ultra-luminous infrared galaxies and two categories of blazars, specifically high-synchrotron-peaked BL Lacs and extreme blazars. The analysis is based on 336 days of data from the Astroparticle Research with Cosmics in the Abyss component of the KM3NeT research infrastructure (KM3NeT/ARCA), collected with detector configurations comprising 19–21 detection units, yielding 1544 selected track-like neutrino candidates. The sample of ultra-luminous infrared galaxies consists of 75 sources, for which conventional power-law spectra are assumed in modelling their signal emission. The blazar samples comprise 232 high–synchrotron-peaked BL Lac objects and 88 extreme synchrotron-peaked emitters, respectively. Their expected neutrino fluxes are obtained from numerical tools constrained by multi-wavelength observational data. No statistically significant excess above the atmospheric background is found for the three catalogues and 90\% confidence level upper limits are set. 
\end{abstract}

\newpage

\section{Introduction} 
\label{sec:intro}
Among astrophysical messengers, high-energy neutrinos provide a unique probe of hadronic processes occurring in extreme astrophysical environments. These neutrinos are thought to originate from the same sources that accelerate cosmic rays, or from their surroundings, where sufficient matter or radiation fields allow part of the energy to be converted into secondary particles~\cite{Kelner:2006tc,Kelner:2008ke}. Because they are neutral and weakly interacting, neutrinos can travel cosmological distances without absorption or deflection, preserving information about their origins. Their detection can, therefore, play a key role in identifying the sources of cosmic rays, even at the highest energies. For this purpose, large-scale neutrino telescopes are being constructed worldwide. The KM3NeT (Cubic Kilometre Neutrino Telescope) Collaboration~\cite{KM3Net:2016zxf,KM3NeT:2024paj}, building on the experience of the ANTARES detector~\cite{ANTLegacy}, is constructing two Cherenkov neutrino detectors in the deep Mediterranean Sea: KM3NeT/ARCA and KM3NeT/ORCA. KM3NeT/ARCA (Astroparticle Research with Cosmics in the Abyss), primarily dedicated to neutrino astronomy, will instrument a volume of about one cubic kilometre and will be sensitive to neutrinos from 100~GeV up to multi-PeV energies. Thanks to its modular design, it is already being collecting data during construction and contributing to high-energy neutrino observations. The KM3NeT/ARCA detector has reported the detection of an ultra-high-energy neutrino with energy of $\sim 220\, \rm PeV$~\cite{KM3NeT:2025npi}, highlighting its capability to identify and reconstruct rare events and demonstrating its scientific potential to probe the most energetic astrophysical processes in the Universe. KM3NeT/ORCA (Oscillation Research with Cosmics in the Abyss), instead, aims at determining the neutrino mass ordering by analysing the flux of atmospheric neutrinos as they propagate through the Earth and studying their oscillation patterns~\cite{KM3NeT:2021ozk}.\\
In this article, a stacked binned likelihood search for high-energy neutrino\footnote{Throughout this paper, the term “neutrino(s)” denotes both neutrinos and antineutrinos ($\nu$ and $\bar{\nu}$).} point sources is presented, based on the information provided by predefined source catalogues. In a stacking analysis, signals from a catalogue of sources are combined to highlight a possible excess from a cumulative flux, thereby increasing the sensitivity of the neutrino telescope. An extended binned likelihood ratio technique is implemented and applied to the case under study. \\
Ultra-luminous infrared galaxies (ULIRG) are objects with infrared luminosities greater than $10^{12}\, L_{\odot}$, with $ L_{\odot}$ being the solar luminosity~\cite{2006asup.book..285L}. Generally, the infrared luminosity is a trademark of stellar formation processes~\cite{Ambrosone:2020evo,Ambrosone:2024xzk,Peretti:2018tmo,IceCube:2021waz}. These sources exhibit dense interstellar media and high supernova explosion rates, which significantly increase the probability of cosmic rays interacting with gas and producing neutrinos via proton–proton collisions. For this reason, ULIRG are well-motivated astrophysical candidates for the production of high-energy neutrinos~\cite{Ambrosone:2024xzk}.\\
Blazars are a subclass of Active Galactic Nuclei (AGN) whose relativistic jets are oriented close to the line of sight to the Earth, resulting in strongly enhanced electromagnetic emission~\cite{Urry:1995mg}. Although blazars constitute a minority of the AGN population, they dominate the extragalactic gamma-ray sky \cite{Fermi-LAT:2019yla}, accounting for roughly 70\% of known sources in the $ \rm GeV-\rm TeV$ energy range. This suggests their efficiency in accelerating cosmic rays to energies exceeding $100\, \rm TeV$~\cite{Blandford:2018iot}. The IceCube Neutrino Observatory has found a correlation at $\sim 3\sigma$ level of high-energy neutrino emission with the blazar TXS-0506+056~\cite{IceCube:2018dnn}. The ANTARES Collaboration has found a $\sim 2\sigma$ correlation between neutrino events and blazars flaring in the radio wavelengths~\cite{ANTARES:2023lck}. Although these results are not yet conclusive, they provide compelling hints of a connection between blazars and neutrino production, strongly motivating further studies. In particular, among the blazar family, high-synchrotron-peaked (HSP) blazars are considered the most probable counterparts of astrophysical neutrinos, being bright and highly variable sources of high-energy gamma rays and candidate accelerators of cosmic rays up to the highest energies~\cite{Pereira:2026sfi}. \\
For ULIRG, sources are chosen from Infrared Astronomical Satellite~\cite{1984ApJ...278L...1N} so that the catalogue is representative of the local distribution of sources as reported by Ref.~\cite{IceCube:2021waz}. HSP blazars are selected from the 3HSP catalogue \cite{Chang:2019vfd}, which currently represents the largest collection of  HSP and extremely HSP blazars, for a total of 2013 candidates. The analysis is split into two source classes, each studied with a stacking approach and different theoretical models for the expected neutrino flux. Specifically, the study considers a subset of HSP blazars analysed with the LeHa-Paris numerical code~\cite{Cerruti:2014iwa}, alongside a set of extreme HSP blazars whose radiative emission is investigated using the LeHaMoC modelling code~\cite{Lehamoc}. Both tools simulate jet emission from supermassive black holes and the particle interactions that can be responsible for the production of high-energy neutrinos. \\
The structure of the paper is as follows: an outline of the KM3NeT detectors is provided in Sec.~\ref{Sec:Detector design and simulation}, along with the associated event simulation and reconstruction procedures. In Sec.~\ref{sec:dataset}, the KM3NeT/ARCA dataset and the event selection strategy adopted in this analysis are described. A summary of the characteristics of the ULIRG and blazar catalogues considered in this work is given in Sec.~\ref{sec:targ}. The likelihood analysis framework is detailed in Sec.~\ref{sec:analysis}. Results from the stacked search for ULIRG and blazars are reported in Secs.~\ref{sec:results_ULIRG} and \ref{sec:disc}, respectively. Conclusions and final discussions are presented in Sec.~\ref{sec:conc}. Finally, in Appendix~\ref{app:A_extrapolation_ULIRG}, the extrapolation of the ULIRG upper limits to the diffuse flux is reported, while in Appendices~\ref{app:B,HBLs} and \ref{app:C,EHBLs} details of the neutrino flux models employed for the two blazar analyses are specified, respectively.

\section{KM3N\lowercase{e}T detectors}
\label{Sec:Detector design and simulation}
\begin{figure}[ht]
    \centering
    \includegraphics[width=0.8\textwidth]{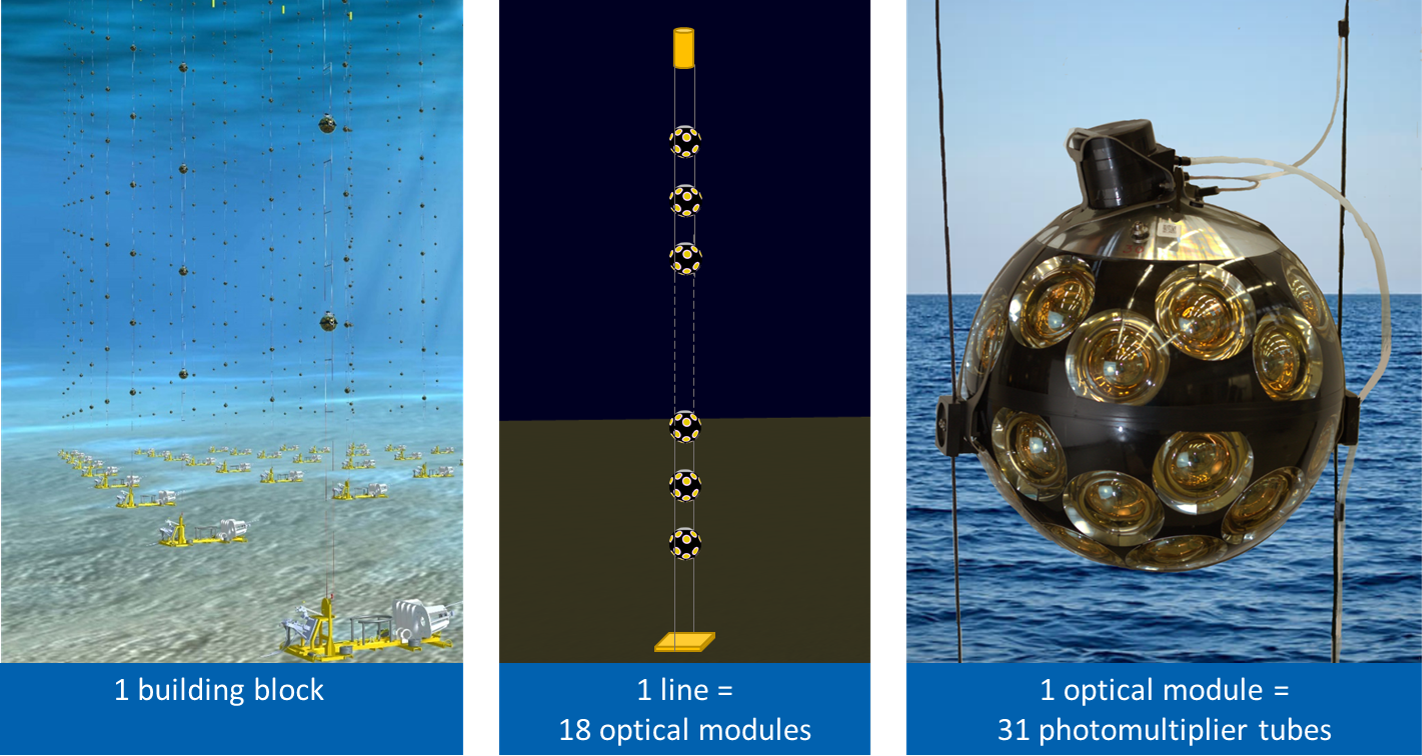}
    \caption{KM3NeT detector layout: KM3NeT in its complete form will consist of 3 building blocks (two for KM3NeT/ARCA and one for KM3NeT/ORCA), each consisting of about a hundred DUs. All DUs support 18 DOMs, which house 31 PMTs each.} 
    \label{fig:detector_layout}
\end{figure}

KM3NeT is a research infrastructure housing two deep-sea Cherenkov neutrino detectors, currently under construction on the seafloor of the Mediterranean Sea. They are designed to study neutrinos across a broad energy range, from a few GeV up to the multi-PeV scale, and to address key questions in both neutrino physics and astrophysics. These include measuring neutrino oscillation parameters, determining the neutrino mass ordering, and investigating the origin of cosmic neutrinos through the detection of astrophysical sources. Thanks to its geographical location in the Northern Hemisphere, KM3NeT provides a field of view that includes the Galactic Centre. The well-understood optical properties of seawater allow for a sub-degree angular resolution for TeV-scale muon neutrinos~\cite{KM3NeT:2024paj}, a key feature for pinpointing point-like or extended sources in the sky.\\
The high-energy detector, KM3NeT/ARCA, is located about 100 km off the Sicilian coast near Portopalo di Capo Passero, Italy. Optimised for TeV–PeV neutrinos, KM3NeT/ARCA aims at detecting and characterising astrophysical neutrino sources and at studying the most extreme cosmic accelerators. KM3NeT/ORCA is situated about 40 km off the coast near Toulon, France. It is optimised for GeV–TeV neutrinos and primarily designed to study neutrino oscillations and neutrino mass ordering. Due to its dense instrumentation, it also enables searches for low-energy astrophysical neutrinos, thereby extending the reach of KM3NeT to softer spectra and complementing KM3NeT/ARCA’s high-energy observations.\\
Upon completion, KM3NeT/ORCA will consist of one building block and KM3NeT/ARCA will comprise two of them, each building block being composed of about a hundred detection units (DUs). Each vertical DU, a flexible string about 700 m high for KM3NeT/ARCA, and 200 m for KM3NeT/ORCA, is equipped with 18 digital optical modules (DOMs). Each DOM is a 0.44 m-diameter pressure-resistant glass sphere capable of withstanding up to $6.7\times10^7$ Pa, housing 31 three-inch photomultiplier tubes (PMTs). Altogether, the PMTs provide a total photocathode area of about 1300 cm$^2$ per DOM \cite{KM32207}. 
In addition to the PMTs, each DOM houses calibration instruments, electronics for power, as well as a Central Logic Board~\cite{KM32310}. Data are transported to the control station on shore via an optical data transport system \cite{KM32302, KM31912, KM32101}. A central software component known as the Control Unit supervises and controls the Data Acquisition (DAQ) system of the detector~\cite{KM32006}. The layout of the KM3NeT detectors is shown in Fig.~\ref{fig:detector_layout}. \\ 
In July 2026, KM3NeT/ARCA consists of 51 DUs, while KM3NeT/ORCA comprises 42 DUs. More details on the deep-sea deployment of the KM3NeT neutrino telescope DUs can be found in Ref.~\cite{KM32011}. 

\subsection{Neutrino detection principle}
Charged particles produced in high-energy neutrino interactions induce the emission of Cherenkov radiation in the surrounding seawater, which is detected with nanosecond timing precision by the PMTs. Each detected photon is converted into an electrical signal via the photoelectric effect. The multi-PMT design~\cite{KM32207}, in contrast to the single large-PMT approach used in ANTARES~\cite{ANTARES:2005hwh}, Baikal-GVD~\cite{Avrorin:2022lyk} and IceCube~\cite{IceCube:2010dpc}, allows intrinsic directional sensitivity to incoming photons and allows better suppression of optical background from $^{40}$K decays and bioluminescence. Each PMT signal, characterised by its leading-edge time, time-over-threshold, and PMT identifier, defines a hit. The number of hits, and their spatial and temporal correlations are used to remove the background of atmospheric muons and to infer the event topology, direction and energy. The topologies (shower or track) seen in the detector are related to the flavour of the interacting neutrino ($\nu_e$, $\nu_\mu$, or $\nu_\tau$) and its interaction types, namely charged current (CC) or neutral current (NC). Shower-like events are mainly created by secondary leptons and hadronic fragmentation in $\nu_{e}$ and $\nu_{\tau}$ CC interactions as well as in NC interactions of all neutrino flavours. In such a case, Cherenkov photons are emitted almost isotropically from a single vertex.\\
The present analysis focuses on track-like events, which predominantly originate from CC interactions of $\nu_\mu$, resulting in a hadronic shower as well as a muon. The muon typically travels long distances (of the order of $\rm km$, at TeV energies). The resulting light trace is well suited for pointing back to the direction of the incoming neutrino. Furthermore, tracks can also be detected when the neutrino interaction vertex is located outside of the detector. This greatly increases the detection efficiency compared to showers. \\
To get from a set of hits in the detector to the physical trajectory and energy of the detected particles, a reconstruction algorithm fits a muon track hypothesis to the associated hit patterns, based on a model of the expected arrival times of Cherenkov photons \cite{Jon23}. For this non-linear problem, an approach with several consecutive steps is employed.
Firstly, different track directions are tested based on the simple hypothesis of a muon track emitting Cherenkov photons under the fixed Cherenkov angle of $42^{\circ}$. The best directions are those that maximise the likelihood in a fit based on analytical Probability Density Functions (PDFs) for the Cherenkov photon arrival times. These PDFs take the scattering of the light into account, as well as radiation from bremsstrahlung showers and delta-rays. The fit selects the best direction as well as the vertex position of the track. A consecutive step then estimates the length of the track based on the first and last point where the PMTs register signal-like hits. Lastly, with a likelihood maximisation based on which PMTs did and which did not record a hit, the energy of the particle is reconstructed. 

\subsection{Sources of background}
Apart of the white noise due to bioluminescence and radioactive decays of $^{40}$K in seawater, the  main sources of background for the search of cosmic neutrinos in deep-sea environments are secondary particles produced in cosmic-ray air showers. Atmospheric muons cannot traverse the Earth, and are expected only as downward-going particles. Therefore, their contribution can be reduced by selecting tracks reconstructed as upgoing. Atmospheric neutrinos represent an irreducible background.  Atmospheric neutrinos have an almost isotropic arrival direction distribution and are expected to have a different (softer) energy spectrum compared to that expected for cosmic neutrinos. Furthermore, the multi-PMT design of the optical modules provides information on the number of incident photons as well as on their arrival times and directions. This information is used in trigger and reconstruction algorithms to highly reduce bioluminescence and $^{40}$K backgrounds \cite{KM32002}. Nevertheless, the light from $^{40}$K decay, and that induced by atmospheric muons, is also useful since it can be used to calibrate the detector elements in time, which complements the calibrations done in addition to measurements in the laboratory~\cite{KM3NeT:2021lsb}.

\subsection{Monte Carlo simulations}
KM3NeT exploits different Monte Carlo generators to simulate neutrino and cosmic-ray interactions, following a run-by-run approach, pioneered by the ANTARES Collaboration~\cite{ANTARES:2020bhr}, that accounts for varying data-taking conditions. Neutrino interactions are generated using the \texttt{gSeaGen} code~\cite{KM3NeT:2020tvi,Gar21}, which interfaces with the \texttt{GENIE} framework~\cite{Andreopoulos:2009rq} for event generation.
\texttt{gSeaGen} simulates neutrino interactions for all flavours over energies from a few~MeV to EeV, using the deep inelastic scattering model CSMS11~\cite{Gar20}. The software also accounts for density and composition of the medium surrounding the detector, ensuring precise modelling of neutrino interactions. \\
The total atmospheric neutrino flux includes both conventional and prompt components.  The conventional flux is modelled using Honda2006~\cite{Hon07}, while the prompt component is based on the ERS model~\cite{Enb08}. Both include an updated determination of the knee in the cosmic-ray spectrum based on the Gaisser-H3a model~\cite{Gai12}. The fluxes are parametrised as a function of the neutrino flavour, energy and direction.
Atmospheric muons and muon bundles from cosmic ray interactions in the atmosphere are simulated using \texttt{MUPAGE}~\cite{Car08}, which parametrises the muon flux in seawater as a function of depth, zenith angle, energy, and, for bundles, multiplicity and radial distance from the bundle axis. These parametrisations are derived from full Monte Carlo simulations of cosmic-ray showers~\cite{Car08,Bec06}. Astrophysical neutrino fluxes can be described either using physically motivated models, derived from numerical simulations, or using simplified benchmark parametrisations. A commonly adopted example of the latter is 
a diffuse cosmic neutrino flux for each (anti-)neutrino flavour $\alpha$, in the form of a $E^{-2.0}$ power law \cite{IceCube:2018fhm}:

\vspace{0.3 cm}

\begin{equation}
\Phi_{\overset{}{\nu_{\alpha}}}^{\rm cos} = 0.6 \times 10^{-8} \left( \frac{E_{\overset{}{\nu_{\alpha}}}}{\rm GeV} \right)^{-2.0} \, \rm GeV^{-1}\,cm^{-2}\,s^{-1}\,sr^{-1}. \label{eq:cosmic_flux} 
\end{equation}

\newpage

\section{Dataset and event selection} 
\label{sec:dataset}
\begin{figure}[h!]
    \centering
    \includegraphics[width=0.6\textwidth]{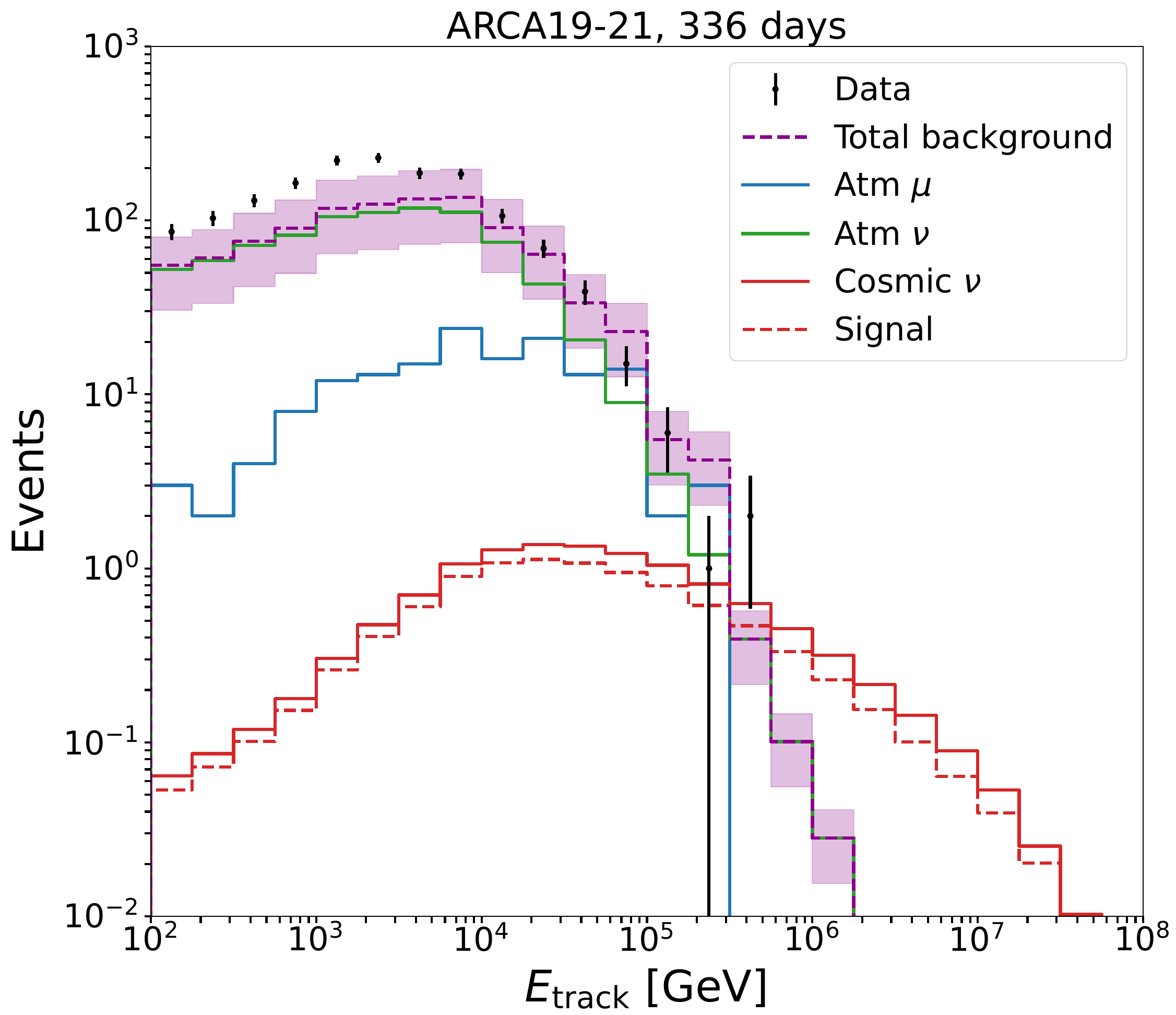}
    \caption{Reconstructed energy distributions after applying the event selections for the ARCA19--21 period. The diffuse cosmic neutrino contribution, assuming the flux in Eq.~\ref{eq:cosmic_flux}, is shown in red, with the dotted line indicating the fraction of $\nu_\mu$ CC track events reconstructed within $1^\circ$ from the true neutrino direction and that produce at least one hit in two different DOMs (labelled as “signal”). Atmospheric muons and neutrinos are shown in green and blue, respectively, while data are shown as black points, together with their statistical uncertainties. The total expected background of atmospheric muons and neutrinos is shown in dashed purple, together with its uncertainty band.}
    \label{fig:zenE_cut2}
\end{figure}

This analysis focuses on data collected with the KM3NeT/ARCA detector in configurations with 19 to 21~(ARCA19-21) operating DUs. The dataset comprises 48.4 days with 19 DUs (ARCA19) and 287.4 days with 21 DUs (ARCA21), totalling a livetime of approximately 336 days. Each dataset is associated with corresponding instrument response functions (e.g. angular resolution, energy response, effective area), which are used to model neutrino signals. The event selection procedure is designed to isolate upgoing, well-reconstructed neutrino-induced muon tracks while suppressing the large background of atmospheric muons and optical noise through successive quality cuts. After the application of preliminary cuts, the reconstructed zenith angle ($\theta_{\text{trk}}$) is required to be $\cos(\theta_{\rm trk})> -0.1$ to reject the overwhelming downgoing muon background among the reconstructed events. Next, noise events (random coincidences, $^{40}$K, and bioluminescence), which typically produce few coincident hits and are hard to reconstruct, are removed by requiring a minimum number of causally-connected hits (selection called “antinoise cuts”). The final selection step includes cuts on the track length, its directional uncertainty, and to the likelihood of the reconstruction. Then, high-purity track-like neutrino candidates are selected using a “track score” provided by a Boosted Decision Tree (BDT) classifier, trained on 20 reconstruction-based variables, such as fit quality, number of hits, track length, and directional uncertainty. In addition, the ultra-high-energy event KM3-230213A~\cite{KM3NeT:2025npi} is removed, since the data are employed to model the background. The inclusion of such an extreme event would introduce a bias in the background estimation. The reconstructed energy distribution of the event samples is shown for the ARCA19-21 data-taking period. Assuming the diffuse cosmic flux of Eq.~\ref{eq:cosmic_flux}, the ARCA19–21 dataset is expected to contain 12 cosmic neutrino events in the energy range $10^2$–$10^8$~GeV. Of these, about 10 events are $\nu_\mu$ CC track events reconstructed within $1^\circ$ of the true neutrino direction. The ARCA19–21 sample is characterised by a residual atmospheric muon contamination of about 15\%, which is expected to decrease further for larger detector configurations. The resulting track energy distributions are shown in Fig.~\ref{fig:zenE_cut2}. A 45\% systematic uncertainty is applied to the total Monte Carlo background prediction to account for the normalisation uncertainties of the atmospheric muon and neutrino fluxes~\cite{Honda:2006qj}, as well as for the detector acceptance (see below for further details). The flux normalisation uncertainty is included for completeness, although it does not affect the analysis presented in this work, whose background estimate is data-driven, as explained in~Sec.\ref{sec:analysis}.

\begin{table}[h!]
\centering
\begin{tabular}{l S S S S}
\toprule
\multicolumn{5}{c}{\textbf{ARCA19--21 - Event statistics}} \\
\midrule
& \textbf{Preliminary cuts} & \textbf{Upgoing} & \textbf{Antinoise} & \textbf{Final selection} \\
\midrule

Data                & \num{1.19e8} & \num{1.28e6} & \num{2.94e5} & \num{1544} \\
Monte Carlo total   & \num{1.16e8} & \num{1.25e6} & \num{3.38e5} & \num{1025} \\
Muons               & \num{1.16e8} & \num{1.24e6} & \num{3.35e5} & \num{150} \\
Atmospheric $\nu$   & \num{5590} & \num{3966} & \num{3223} & \num{863} \\
Cosmic $\nu$        & \num{66} & \num{36} & \num{32} & \num{12} \\
Signal & \num{25} & \num{14} & \num{12} & \num{10} \\

\bottomrule
\end{tabular}
\caption{Summary of the event statistics at different selection steps for the ARCA19--21 detector. Monte Carlo cosmic contributions are weighted with the expression reported in Eq.~\ref{eq:cosmic_flux}. The label “signal” refers to $\nu_\mu$ CC track events reconstructed within $1^\circ$ of the true neutrino direction and that produce at least one hit in two different DOMs.}
\label{tab:event_stats}
\end{table}

\newpage

\subsection{Background expectation from scrambled data}
\label{background expectation}

In order to evaluate the sensitivity for each analysed detector configuration, pseudo-experiments are generated simulating background and signal. Due to the Earth's rotation and the fact that the detector is steadily in operation, background events are randomly generated by sampling the declination ($\delta$) from data and scrambling their arrival times. This corresponds to sampling the right ascension (R.A.) from a uniform distribution, while the declination distribution depends on the detector acceptance. The dependencies of the background rates in the declination and the reconstructed energy are assumed to be factorisable. The reconstructed energy component is modelled by fitting the scrambled dataset with a sum of Gaussian functions. The declination component is modelled using a spline interpolation. A normalisation factor is applied to ensure that the total number of background events integrated over the full solid angle sky corresponds to the number of events in the total dataset for the evaluated period. 

\subsection{Detector response from Monte Carlo simulations}
\label{detector response}
\begin{figure}[h]
    \centering
    \includegraphics[width=0.46\textwidth]{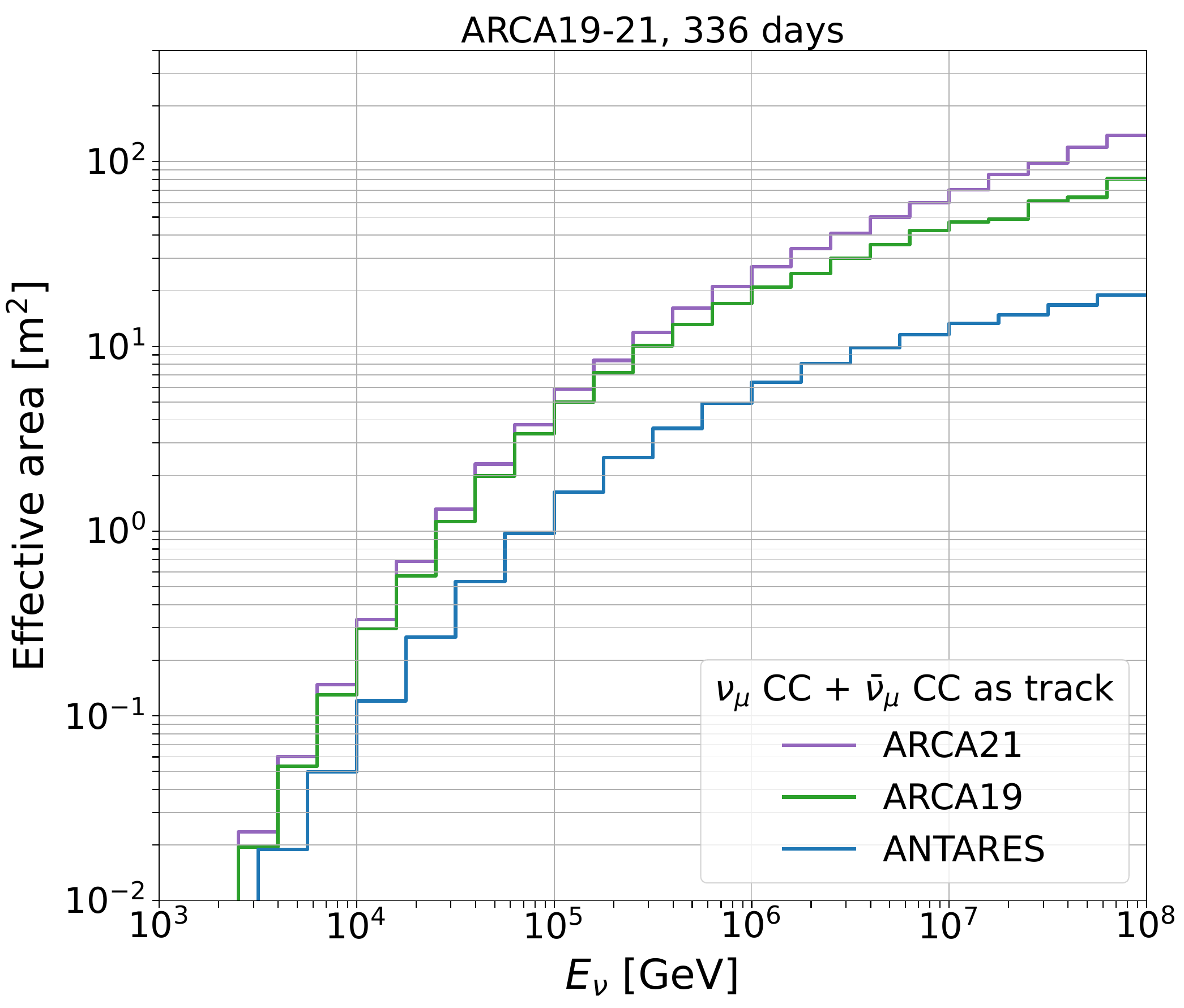}
    \hspace{0.014\textwidth}
    \includegraphics[width=0.46\textwidth]{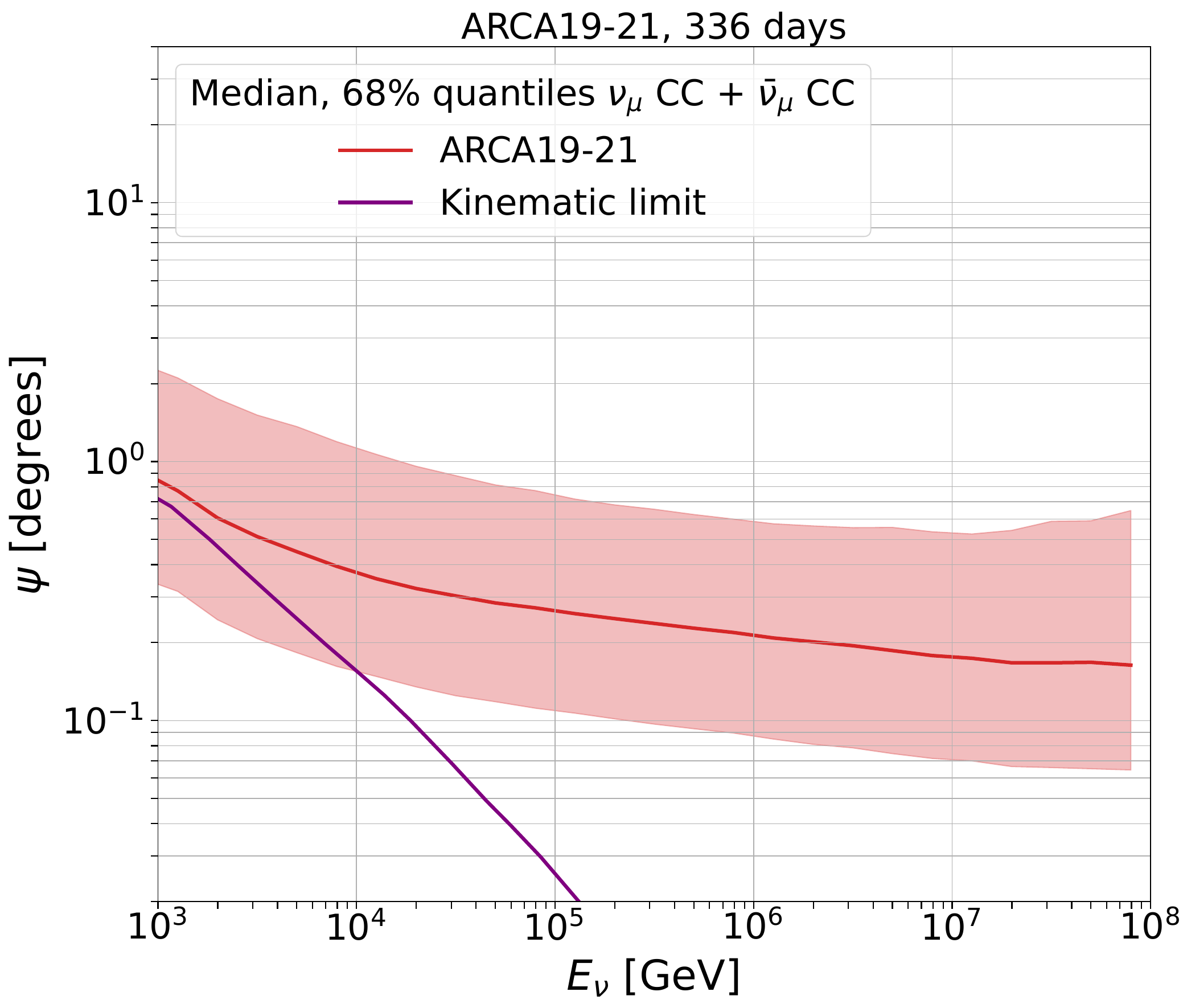}
    \caption{\textbf{Left:} Effective area as a function of the true neutrino energy for different KM3NeT/ARCA configurations, averaged between neutrinos and antineutrinos. The effective area of ANTARES is shown for comparison~\cite{ref:192}. \textbf{Right:} Median angular deviation ($\psi$) between the reconstructed and true neutrino direction as a function of the true neutrino energy, together with the 68\% quantiles represented by the shaded area. Both plots show the detector performance for $\nu_\mu^{\rm CC} + \bar{\nu}_\mu^{\rm CC}$ at the final selection level described in Sec.~\ref{sec:dataset}. The kinematic angle between the neutrino and the outgoing lepton, in CC interactions, is shown as a purple line.}
    \label{fig:EffArea}
\end{figure}

To determine the signal expectation at a given sky location, the response of each detector configuration to a possible signal is modelled as a function of energy, zenith and azimuth, based on the earlier discussed simulations.  The angular and energy resolutions, obtained by comparing the true and reconstructed values for each simulated event, are used to compute the expected spread of events around a source location in energy. The effective area is used to compute the expected number of events at a given declination and flux. In Fig.~\ref{fig:EffArea} the effective area of the different detector configurations is shown, as well as the angular resolution for selected track events from $\nu_\mu^{\rm CC}$ and $\bar \nu_\mu^{\rm CC}$ interactions. The sky-averaged effective area of ANTARES is also reported \cite{ref:192}, although the respective data samples have different muon contamination levels (14\% for the ANTARES track-like event selection). The estimated angular deviation at 1 PeV is 0.15$^\circ$.

\section{Target sources}
\label{sec:targ}
\begin{figure}[h!]
    \centering
    \includegraphics[width=0.75\linewidth]{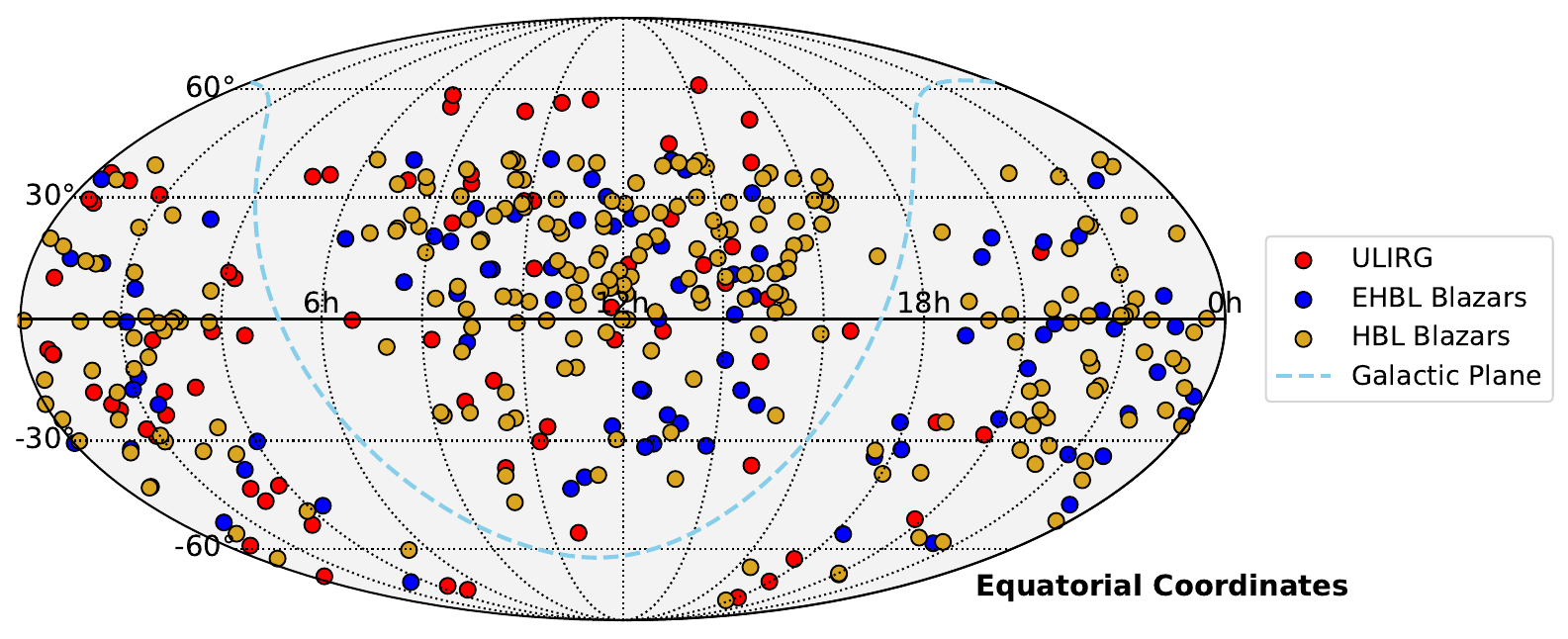}
    \caption{Sky map, in equatorial coordinates, of the sources analysed in this study: ULIRG (red dots), high–synchrotron-peaked blazars (orange dots) and extreme high–synchrotron-peaked BL Lacs (blue dots). The Galactic plane is marked by a light-blue dashed line.}
    \label{fig:skymap}
\end{figure}

The ULIRG catalogue analysed in this work is the one reported in Ref.~\cite{IceCube:2021waz}. It comprises 75 local sources with luminosity distances between $80$ and $630\, \rm Mpc$. The sources have been selected from three different catalogues extracted from the Infrared Astronomical Satellite (IRAS) data, with  a threshold of $\sim 1$ $\rm Jy$ for the measured flux density at $60\, \rm \mu m$.  This catalogue is representative of the local ULIRG population up to redshift $z \simeq 0.13$~\cite{IceCube:2021waz}, pertaining to the observational distance threshold of the flux density at $60\, \rm \mu m$~\cite{IceCube:2021waz}. No further declination cut is considered to ensure fair comparison with IceCube results~\cite{IceCube:2021waz} and to allow the extrapolation of the results to the entire source population~(See Appendix~\ref{app:A_extrapolation_ULIRG} for further details).
For blazars, only BL Lacertae (BL Lacs) objects are selected, which are characterised by a featureless non-thermal continuum in the optical/UV spectrum. The selection is made according to the following requirements: a secure source identification and a reliable redshift measurement, as flagged in the 3HSP catalogue, and a declination below $40^\circ$ to account for the visibility of KM3NeT/ARCA~\cite{KM3Net:2016zxf}. The synchrotron peak frequency, $\nu_{\text{peak}}^{\text{syn}}$, defined as the frequency at which the synchrotron component of the blazar spectral energy distribution (SED) reaches its maximum, is commonly used to classify BL Lac objects. Sources with $10^{15.0}\,\text{Hz} < \nu_{\text{peak}}^{\text{syn}} < 10^{16.8}\,\text{Hz}$ are classified as high–synchrotron–peaked BL Lacs (HBL), while sources with $\nu_{\text{peak}}^{\text{syn}} \geq 10^{16.8}\,\text{Hz}$ are classified as extremely high–synchrotron–peaked BL Lacs (EHBL). For the EHBL sample, an additional requirement is imposed, selecting only the 3HSP sources that present a $\gamma$-ray counterpart in the Fermi–LAT 4FGL catalogue \cite{Fermi-LAT:2019yla}. This finally results in the selection of 232 objects for the HBL catalogue and 88 objects for the EHBL catalogue.\\ The spatial distribution of all selected sources in equatorial coordinates is presented in Fig.~\ref{fig:skymap}.

\noindent

\section{Analysis framework} 
\label{sec:analysis}
\begin{figure}[h!]
    \centering
    \includegraphics[width=0.95\linewidth]{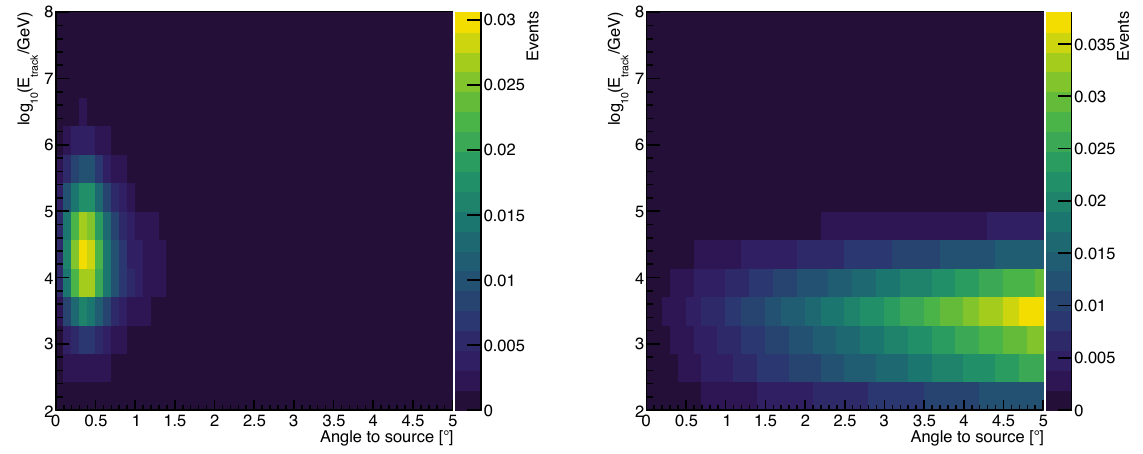}
    \caption{\textbf{Left:} Expected signal PDF at declination $\delta$=$-85^\circ$ for ARCA21, assuming an astrophysical neutrino spectrum following an $E^{-2.0}$ power-law. As it can be seen, the signal clusters near the source. \textbf{Right:} Expected background PDF at declination $\delta$=$-85^\circ$ for ARCA21. The rate of background dominates in bins covering larger solid angles.}
    \label{fig:KM3NeT/ARCA21PDFbin}
\end{figure}

First, the neutrino fluxes from the individual sources of each catalogue are estimated. For the ULIRG case, the signal flux is assumed to be a power-law flux $\phi(E) = \phi_0 E^{-\gamma}$ (see more details in Appendix~\ref{app:A_extrapolation_ULIRG}), fixing the value of the spectral index $\gamma$ to several tested values, while for blazars the shape of the expected neutrino flux is estimated following the methods described in Appendices~\ref{app:B,HBLs}, in the case of HBL, and \ref{app:C,EHBLs}, in the case of EHBL. Flux predictions have been optimised by fitting multi-wavelength photon data with numerical codes that explore the blazar parameter space. The signal and background expectations, at the final selection level, for declination $\delta = -85^{\circ}$ and 21 DUs, are shown in Fig.~\ref{fig:KM3NeT/ARCA21PDFbin}, assuming an $E^{-2.0}$ spectrum.  Second, an extended binned maximum likelihood approach is adopted, building two-dimensional distributions of the observed event rate in two-dimensional histograms in terms of the reconstructed track energy, in the range [2;8] in $\log_{10}({E_{\rm track}/GeV})$, and the angular distance within an angle to the source $\alpha = 5^{\circ}$~\cite{KM3NeT:2024paj,KM3NeT:2024uhg}. The signal is concentrated near the source and generally occurs at higher energies, with a spectrum that is generally harder than that of the background. The reason why signal events appear in the leftmost bins is that the bins represent increasing solid angles from left to right. Consequently, the rightmost bin, which covers the largest solid angle, contains the majority of background events. 
The background distribution is obtained using scrambled data as described in Section 3.1. The signal distribution $S_i$, instead, is evaluated by means of Monte Carlo simulations and depends on the shape of the assumed flux. 
The effect of systematic uncertainties on the PDFs is also considered. The log-likelihood function for a given dataset is expressed as:
\begin{equation}
\log \mathcal{L}(\zeta) = \sum_{i \in \text{bins}} N_i \log(B_i + \zeta S_i) - B_i - \zeta S_i\, ,
\end{equation}
where $\zeta$ is the signal strength (a free normalisation parameter for the reference flux) and $N_i$ is the observed number of events. The optimal signal strength $\hat{\zeta}$ is determined by maximising $\log \mathcal{L}$. The test statistic $\lambda$ is defined as the logarithm of the likelihood ratio:
\begin{equation}\label{eq:TS}
\lambda = \log \left( \frac{\mathcal{L}(\zeta = \hat{\zeta})}{\mathcal{L}(\zeta = 0)} \right),
\end{equation}
and is used to quantify the preference for the signal+background hypothesis (H1) over the background-only one (H0). The test-statistic distributions are estimated using pseudo-experiments (PEs). In particular, for a given value of $\zeta_{\text{true}}$, the corresponding $\lambda$ distribution is determined by randomly generating events from a Poisson distribution ($P$) with a mean equal to $B_i + \zeta_{\text{true}} S_i$. For $\zeta_{\text{true}} =0$, $10^5$ PEs are generated, while $2 \times 10^4$ PEs are simulated for the other cases.
The model rejection factor, referred to as $\zeta_{90}$,  is defined as in Ref.~\cite{KM3NeT:2024uhg}:
\begin{equation}\label{eq:MRF}
    \int^{+\infty}_{\rm \lambda_{m}} \left. \frac{dP}{d\lambda} \right|_{\rm H1(\zeta = \zeta_{90})} d\rm \lambda = 90\%,
\end{equation}
where ${\rm \lambda_{m}}$ is the median of the distribution for the background-only hypothesis. The sensitivity is defined as $\phi_{90}(E) = \zeta_{90} \phi_s (E)$ where $\phi_s (E)$ is the total flux injected corresponding to $\zeta =1$. By substituting $\lambda_{\rm m}$ with $\lambda_{\rm obs}$, namely the observed $\lambda$ on the unblinded dataset, the 90\% C.L. upper limits $\zeta_{\rm UL}$ are obtained according to the relation:
\begin{equation}\label{eq:uL}
    \int^{+\infty}_{\rm \lambda_{\rm obs}} \left. \frac{dP}{d\lambda} \right|_{\rm H1(\zeta = \zeta_{\rm UL})} d\rm \lambda = 90\%.
\end{equation}
To increase sensitivity to weak signals across multiple sources, a stacking analysis is implemented. The source candidates are ranked by their expected neutrino yield given by the assumed spectral model and their likelihood profiles are combined according to:
\begin{equation}\label{eq:stacking_likelihood}
\log \mathcal{L}_{\text{combined}}(\zeta) = \sum_j \log \mathcal{L}_j(\zeta),
\end{equation}
where $j$ indexes the sources. This approach preserves the relative contribution of each source and allows for a single signal strength parameter $\zeta$ to be fitted globally. For the stacking analysis, the sensitivity  is defined with the  frequentist PE's approach with Eqs.~\ref{eq:TS}-\ref{eq:MRF} as used in a standard point-source analysis.

\subsection{Systematic uncertainties}
The performance results are evaluated including systematic effects. For KM3NeT/ARCA, the dominant source of angular uncertainty originates from a potential misalignment of the detector’s absolute orientation around the vertical axis, i.e. the position of the telescope’s mechanical structures, in a geo-referenced coordinate system. The relative positions of the optical modules are known with a precision better than $20 \, \rm cm$, using acoustic emitters/receptors on the modules and autonomous beacons to monitor in real time their relative distances, tilts, and rotations. However, the dominant source of uncertainty on the angular pointing is related to the absolute orientation of the whole system. It is estimated by propagating the uncertainties on the positions of the acoustic beacons, as well as accounting for potential tilts due to slopes on the seabed. An independent cross-check was performed by considering the deficit of cosmic rays in the direction of the Moon due to their absorption, leading to uncertainties of $0.49^{\circ}$~\cite{aiello2023first} and $0.24^{\circ}$~\cite{KM3NeT:2025npi}, respectively. In the following, a conservative Gaussian error of $0.5^{\circ}$ is adopted to account for the pointing accuracy of both detectors. \\
In addition, systematic effects influence the overall detector acceptance, mainly due to uncertainties in water optical properties and the PMTs response. Dedicated simulations in which these parameters are independently varied indicate a conservative uncertainty at the $\sim$30\% level when all contributions are summed. Further calibration and simulation work is underway to refine these estimates.

\section{Results for ULIRG}
\label{sec:results_ULIRG}
\begin{figure}[h!]
    \centering
    \includegraphics[width=0.495\linewidth]{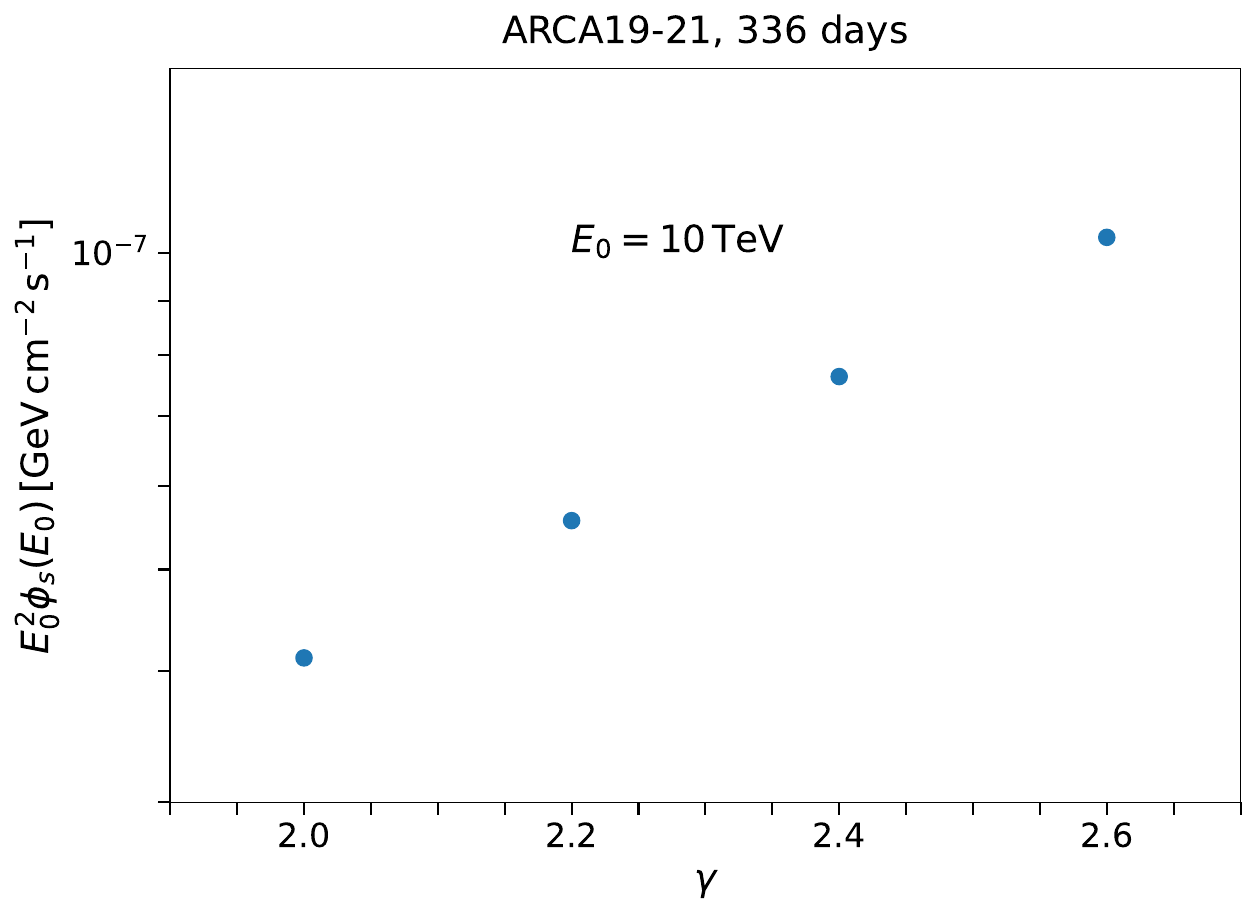}
    \includegraphics[width=0.495\linewidth]{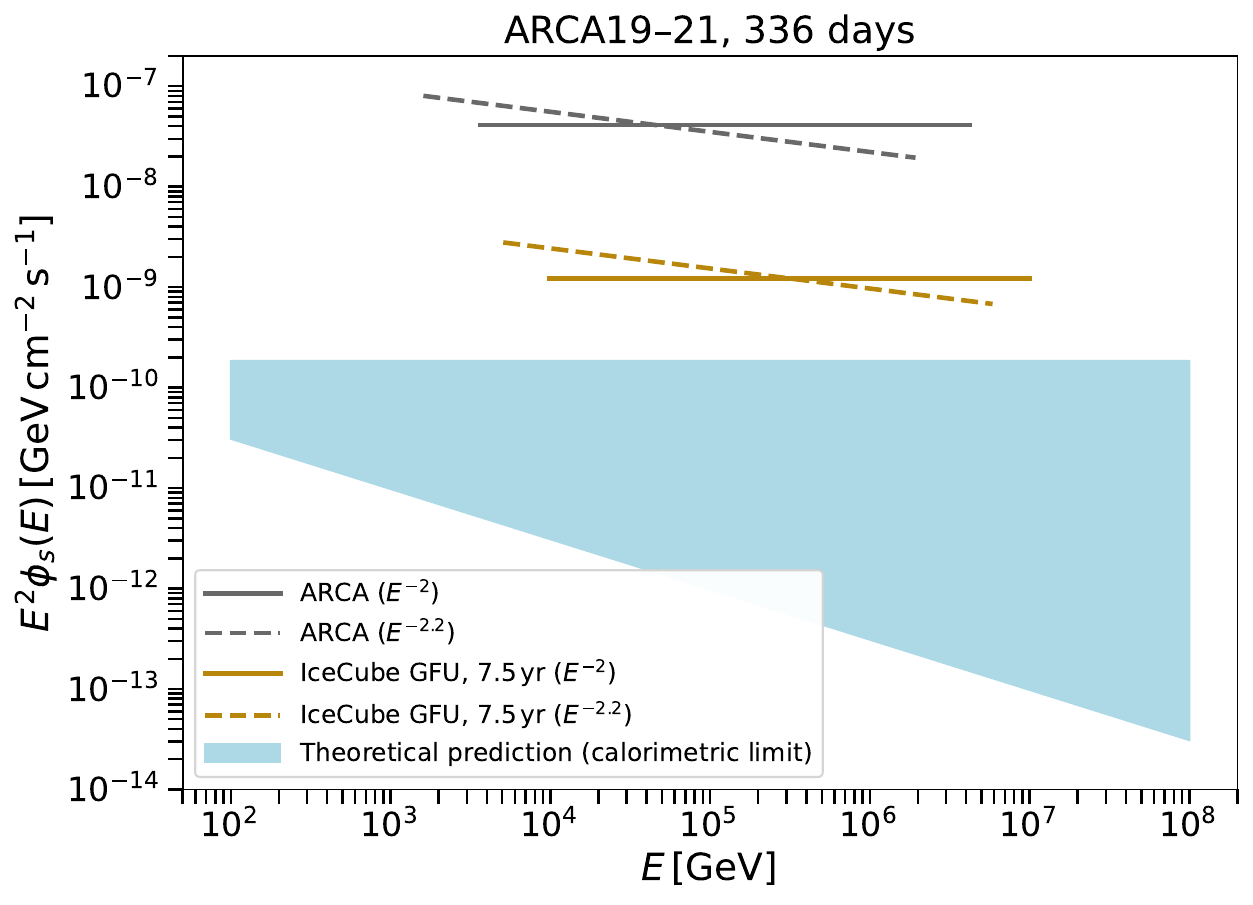}
    \caption{\textbf{Left}: The $90\%$ C.L. upper limit at $E_0 = 10\, \rm TeV$ for the stacking analysis in terms of the spectral index. \textbf{Right}: $E^{-2.0}$ and $E^{-2.2}$ upper limits for ARCA19-21~(grey lines)  and the IceCube~\cite{IceCube:2021waz}~(golden lines) detectors as function of the energy. The upper limits are compared with the theoretical prediction (blue band), evaluated using the calorimetric assumption explained in Ref.~\cite{Ambrosone:2024xzk}. }
    \label{fig:results_1}
\end{figure}

In order to evaluate neutrino emission from star-forming processes for ULIRG, each source is assumed to share the same power-law $E^{-\gamma}$ spectrum. The relative contribution of each source is then encoded through weights assigned to their spectra according to
\begin{equation}
    w_j = \frac{F_{IR}^j}{\sum_{j'} F_{IR}^{j'}} \, ,
\end{equation}
with $F_{IR} = L_{IR}/(4\pi D^2)$, where $L_{IR}$ and $D$ are the infrared luminosity and distance of each source $j$ to the Earth, respectively. This choice is motivated by the fact that infrared luminosity is commonly used as a tracer of the star-formation rate, which is expected to correlate with the acceleration of high-energy cosmic rays~\cite{Ambrosone:2024xzk}.
The analysis yields a best-fit signal strength $\hat{\zeta}\simeq 0$ for $\gamma = 2.0$, compatible with the background-only hypothesis, resulting in a p-value of $0.2$ and a corresponding one-sided significance of $0.9\sigma$.\\
The results of the stacking analysis throughout the entire catalogue are summarised in Fig.~\ref{fig:results_1}. In the left panel, the $90\%$ C.L. upper limits are shown as a function of the assumed spectral index, for a fixed reference energy of $E_0 = 10\,\mathrm{TeV}$.
In the right panel the KM3NeT/ARCA upper limits for $E^{-2.0}$ and $E^{-2.2}$ spectra are compared with the IceCube results~\cite{IceCube:2021waz}, shown as a function of energy within the  90\% central energy range. Also indicated are the theoretical predictions, obtained assuming that the entire infrared luminosity of sources is ascribed to stellar-forming processes and that all the cosmic rays injected produce neutrinos, referred to as the calorimetric assumption (see~Ref.~\cite{Ambrosone:2024xzk}~for more details). Therefore, the prediction can be interpreted as a theoretical upper limit on the neutrino emission for the sources in the catalogue.  Given that the theoretical predictions lie below the observed $90\%$ C.L. upper limits, the absence of detected signal events from ULIRG is compatible with the calorimetric assumption for these sources, as already probed by IceCube. The derived upper limit can be used to constrain the neutrino emission of the whole ULIRG population in the Universe, as detailed in Appendix~\ref{app:A_extrapolation_ULIRG}.

\section{Results for blazars}
\label{sec:disc}

\subsection{Sensitivity and upper limits for the HBL catalogue}

\begin{figure}[h!]
    \centering
    \includegraphics[width=0.65\linewidth]{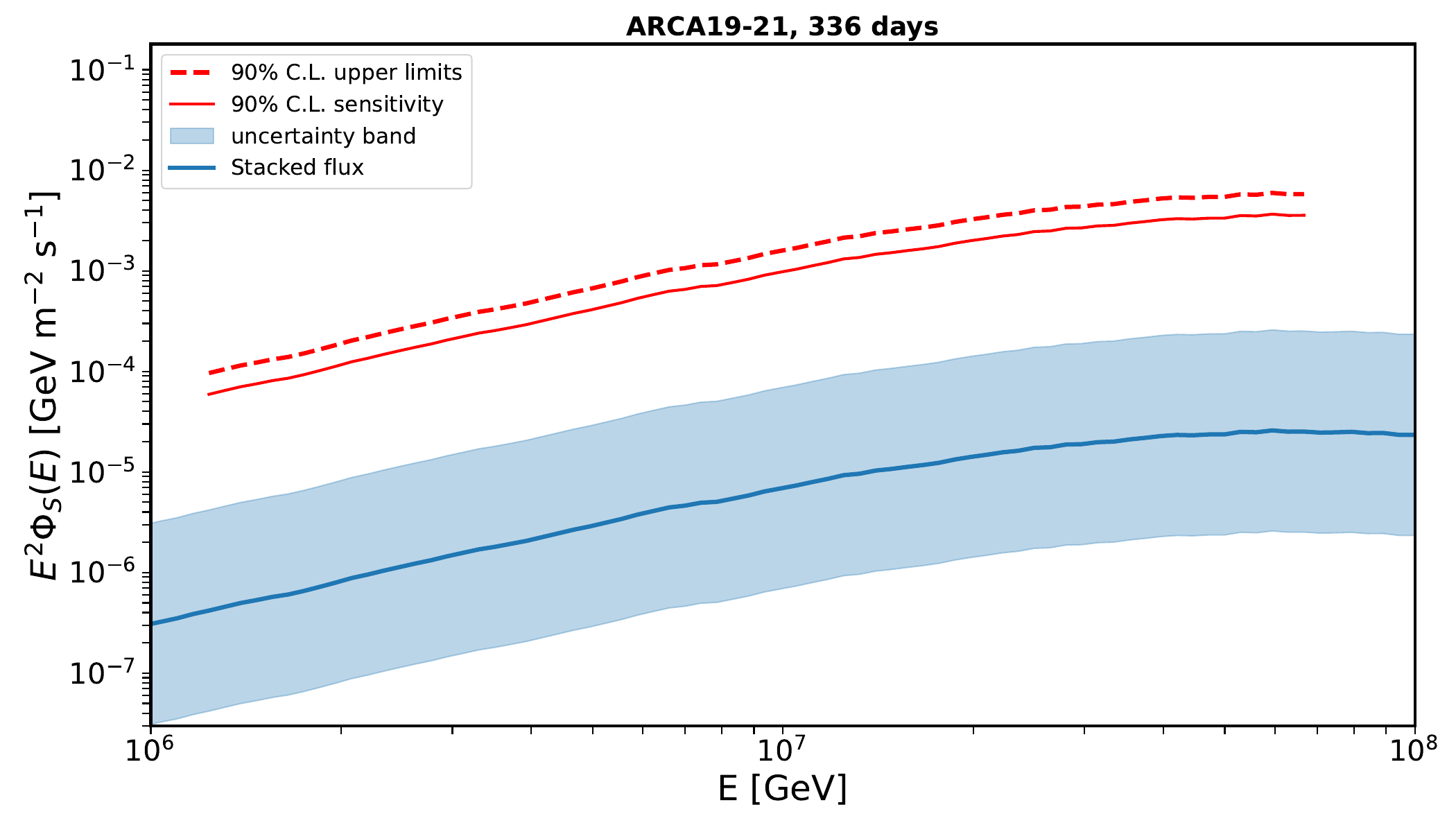}
    \caption{90\% C.L. upper limits and sensitivity~(dashed and solid red line) for the stacked flux (solid blue) expected from the selected HBL sources, shown in the central $90\%$ energy range. The light-blue region indicates the predicted flux uncertainty according to Ref.~\cite{Carenini:2025Gv}.}
    \label{fig:UL_BHL}
\end{figure}

The neutrino emission from HBL is modelled following the approach presented in Ref.~\cite{Carenini:2025Gv}, which considers a single-zone scenario where electrons and protons are accelerated within the relativistic jet of the blazar. The neutrino fluxes for the sources of the catalogue are derived from a reference flux obtained by fitting the multi-wavelength emission of the PKS~2155-304 blazar~\cite{Madejski:2016evb, HESS:2019fhy} using the LeHa-Paris code. This reference flux is then extrapolated to other sources in the catalogue, inferring the neutrino flux normalisation from the synchrotron peak flux and adjusting the neutrino energy spectrum with the source redshift. Therefore, the relative contribution of each source $j$ is determined by the weight
\begin{equation}
w_j = \frac{L_{\gamma,\mathrm{peak}}^j}{L_{\gamma,\mathrm{peak}}^{\mathrm{PKS}}},
\end{equation}
where $L_{\gamma,\mathrm{peak}}^j$ is the synchrotron peak luminosity of the $j$-th source in the catalogue and $L_{\gamma,\mathrm{peak}}^{\mathrm{PKS}}$ is the synchrotron peak luminosity of PKS~2155-304 (see Appendix~\ref{app:B,HBLs} for details). The analysis is performed by progressively adding sources from the catalogue, ordered by brightness, to evaluate how the sensitivity improves with an increasing number of stacked blazars. After the first 100 sources, the incremental gain from adding additional objects becomes negligible; therefore, these 100 sources are selected for unblinding. The analysis yields a fitted signal strength $\hat{\zeta}\simeq 0.35$, indicating no excess over the background-only hypothesis. The observed p-value is $0.04$, corresponding to a statistical significance of $1.7\sigma$, in a one-sided convention. Consequently, the 90\% C.L. upper limit on the stacked flux is derived and shown in Fig.~\ref{fig:UL_BHL} as a function of energy, within the 90\% central energy range, corresponding to approximately $[6.1,7.8]$ in $\log_{10}({E/\rm GeV})$. Fig.~\ref{fig:UL_BHL} also displays the corresponding sensitivity, likewise expressed as a function of energy over the same range. The observed upper limit lies a factor 60\% above the sensitivity. Following the phenomenological approach of Ref.~\cite{Carenini:2025Gv}, an uncertainty band spanning one order of magnitude around the estimated flux is adopted for each source and propagated to their stacked contribution. This uncertainty is estimated using the HBL Mrk~421 \cite{ref:169} and VER~J0521+211~\cite{ref:170} as target sources. These two objects dominate the stacking contribution due to their elevated synchrotron peak fluxes, which strongly regulate the predicted neutrino flux, given that the synchrotron radiation serves as target in lepto-hadronic interactions for producing neutrinos.\\

\subsection{Sensitivity and upper limits for the EHBL catalogue}
\begin{figure}[h!]
    \centering
    \includegraphics[width=0.65\linewidth]{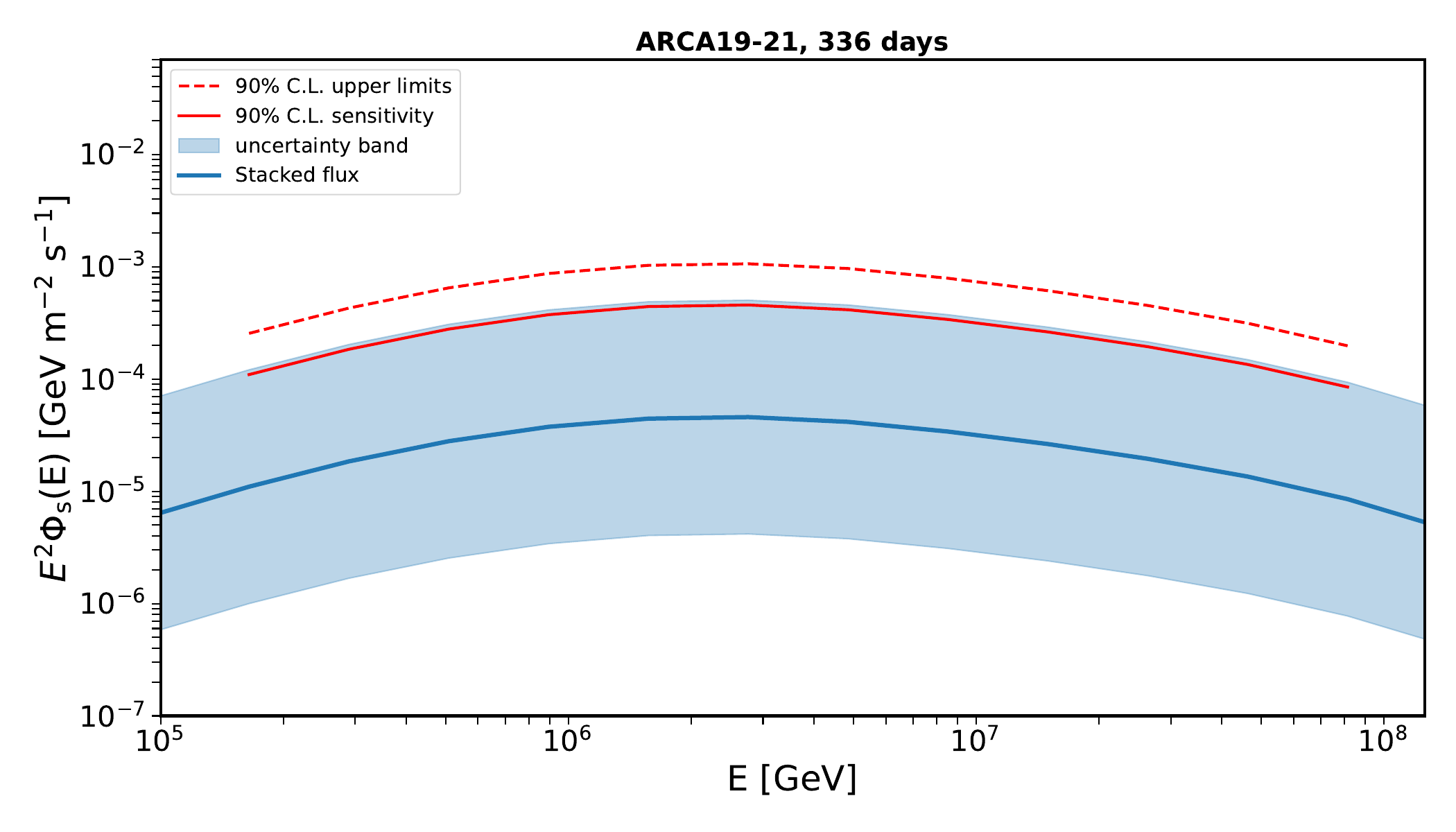}
    \caption{90\% C.L. upper limit and sensitivity~(dashed and continuous red lines) expected from the selected EHBL sources, shown in the central $90\%$ energy range. The shaded light-blue region indicates the predicted flux uncertainty, explained in Appendix~\ref{app:C,EHBLs}.}
    \label{fig:EHBLuL}
\end{figure}

For EHBL sources, the neutrino emission is predicted adopting the open-source lepto-hadronic code LeHaMoC~\cite{Lehamoc}, which considers a one-zone scenario where electrons and protons are accelerated in a blob moving at relativistic speed. The parameters of the sources are determined by fitting the photon data with  a pure electron population. The hadronic component is added by setting the  same shape and parameters both for proton and electron distributions, setting the baryonic loading (the ratio between the proton and electron luminosity) to $10^4$, motivated by the recent analysis detailed in Ref.~\cite{Rodrigues2024}. In the case of EHBL, the relative contribution of each source $j$ is determined by the weight
\begin{equation}
w_j = \frac{F_{j,\gamma}}{F_{\gamma}^{*}},
\end{equation}
where $F_{j,\gamma}$ is the integrated gamma-ray flux of the $j$-th selected source in the $50\,\mathrm{MeV}-1\,\mathrm{TeV}$ Fermi-LAT energy range, and $F_{\gamma}^{*}$ is the corresponding integrated gamma-ray flux of the prototype blazar used to model its neutrino emission. This approach allows for the application of a consistent and homogeneous scaling factor to all sources, ensuring that the scaling flux is measured by the same gamma-ray detector (see Appendix~\ref{app:C,EHBLs} for details). The analysis procedure is analogous to that adopted for HBL, yielding a fitted signal strength $\hat{\zeta}=0.14$, which indicates no significant excess over the background-only hypothesis. The observed $\rm p$-$\rm value$ is $4.0 \times 10^{-3}$, corresponding to a statistical significance of 2.6$\sigma$, in a one-sided convention. The significance is driven by a single source, 3HSP~J204008.3-711459, which represents one of the brightest sources in the catalogue. In Fig.~\ref{fig:EHBLuL} the 90\% sensitivity and upper limit are reported in terms of the energy in the central 90\% energy range, corresponding to $[5.2,7.9]$ in $\log_{10}({E/\rm GeV})$. The sensitivity is approximately one order of magnitude above the stacked mean flux, whereas the upper limit is about a factor 20 above it.  The related flux uncertainty is also shown. \\
It is noteworthy that the use of different numerical frameworks does not change the results of this analysis. In fact, Ref.~\cite{Cerruti:2024lmj} has demonstrated that several numerical codes, e.g. AM$^3$, B13, ATHE$\nu$A, LeHa-Paris and LeHaMoC, provide consistent fluxes within  a $40\%$ difference in normalisation, which cannot modify neither the sensitivities nor the unblinded upper limits presented in this study. \\
To evaluate the post-trial significance of the observed excess from the catalogue of EHBL, 15000 pseudo-skymap realisations of the data are produced. For each realisation, the full stacking analysis is repeated independently for the HBL, EHBL, and ULIRG catalogues, and the lowest p-value across catalogues is recorded, producing the distribution for the minimum p-values expected. By comparing the actual observed significance of 2.6$\sigma$ with this distribution, the global post-trial significance is derived by integrating the distribution above this threshold. This leads to an overall post-trial correction of 1.8$\sigma$, adopting the one-sided convention.

\section{Conclusions}
\label{sec:conc}
The results of a stacking point-source likelihood search with the KM3NeT/ARCA detector with 19 and 21 detection units (data collected from June 2022 to September 2023, for a total livetime of 336 days), using all-flavour neutrino candidates reconstructed as tracks, have been presented. The analysis relies on an $E^{-\gamma}$ power-law assumption for the signal shape in the case of ULIRG, while for HBL and EHBL it employs physically-motivated models derived with numerical codes that simulate the non-thermal radiation processes in blazar jets. After unblinding, none of the catalogues shows a significant excess, and corresponding upper limits have been derived. However, in the case of EHBL, the current upper limit is approaching the range of fluxes predicted by the theoretical model. The ongoing expansion of the KM3NeT/ARCA deep-sea infrastructure, combined with future analyses exploiting increased exposure, larger neutrino statistics, and improved energy and angular resolution, is expected to enhance the sensitivity of these searches.

\newpage

\section*{Acknowledgements}

The authors acknowledge M. Petropoulou and M. Cerruti for their support in the implementation of high-energy neutrino emission from blazars with the LeHaMoc~\cite{Lehamoc} and LeHa-Paris~\cite{Cerruti:2014iwa} numerical codes, respectively. 
The authors acknowledge the financial support of:

%INFRADEV
\noindent KM3NeT-INFRADEV2 project, funded by the European Union Horizon Europe Research and Innovation Programme under grant agreement No 101079679;
%Belgium
Funds for Scientific Research (FRS-FNRS), Francqui foundation, BAEF foundation.
%Czeck
Czech Science Foundation (GAČR 24-12702S);
%France
Agence Nationale de la Recherche (contract ANR-15-CE31-0020), Centre National de la Recherche Scientifique (CNRS), Commission Europ\'eenne (FEDER fund and Marie Curie Program), LabEx UnivEarthS (ANR-10-LABX-0023 and ANR-18-IDEX-0001), Paris \^Ile-de-France Region, Normandy Region (Alpha, Blue-waves and Neptune), France,
%For the CPER
The Provence-Alpes-Côte d'Azur Delegation for Research and Innovation (DRARI), the Provence-Alpes-Côte d'Azur region, the Bouches-du-Rhône Departmental Council, the Metropolis of Aix-Marseille Provence and the City of Marseille through the CPER 2021-2027 NEUMED project,
%For IN2P3
The CNRS Institut National de Physique Nucléaire et de Physique des Particules (IN2P3),
%for ACME
The ACME project funded by the European Union’s Horizon Europe Research and innovation programme under Grant Agreement No 101131928;
%Germany (Max Planck Inst.)
ERC MuSES project No 101142396); ERC starting grant MessMapp, under contract No. 949555.
%Greece
The General Secretariat of Research and Innovation (GSRI), Greece;
%Italy
Istituto Nazionale di Fisica Nucleare (INFN) and Ministero dell’Universit{\`a} e della Ricerca (MUR). KM3NeT4RR MUR Project National Recovery and Resilience Plan (NRRP), Mission 4 Component 2 Investment 3.1, Funded by the European Union – NextGenerationEU,CUP I57G21000040001, Concession Decree MUR No. n. Prot. 123 del 21/06/2022;
%Morocco
Ministry of Higher Education, Scientific Research and Innovation, Morocco, and the Arab Fund for Economic and Social Development, Kuwait;
%The Netherlands
Nederlandse organisatie voor Wetenschappelijk Onderzoek (NWO), the Netherlands;
%Poland
The grant “AstroCeNT: Particle Astrophysics Science and Technology Centre”, carried out within the International Research Agendas programme of the Foundation for Polish Science financed by the European Union under the European Regional Development Fund; The program: “Excellence initiative-research university” for the AGH University in Krakow; The ARTIQ project: UMO-2021/01/2/ST6/00004 and ARTIQ/0004/2021;
%Romania
Ministry of Education and Scientific Research, Romania
%Slovak Republic
Slovak Research and Development Agency under Contract No. APVV-22-0413; Ministry of Education, Research, Development and Youth of the Slovak Republic;
%Spain
MICIU for PID2024-156285NB-C41, -C42- C43, funded by MICIU/AEI/10.13039/501100011033 and by FEDER, EU, and for CNS2023-144099; Generalitat Valenciana for CIDEGENT/2020/049, CIDEGENT/2021/23, CIDEIG/2023/20, CIPROM/2023/51 and INNVA1/2024/110 (IVACE+i), and Fundaci\'{o}n Bancaria La Caixa (ID 100010434), for LCF/BQ/PI25/12100025, Spain;
%UAE
Khalifa University internal grants (ESIG-2023-008, RIG-2023-070 and RIG-2024-047), United Arab Emirates;
%UK
The European Union's Horizon 2020 Research and Innovation Programme (ChETEC-INFRA - Project no. 101008324).
% disclaimer
Views and opinions expressed are those of the author(s) only and do not necessarily reflect those of the European Union or the European Research Council. Neither the European Union nor the granting authority can be held responsible for them.

\newpage
\appendix

\section{ULIRG stacking analysis: extrapolation to the whole sky}\label{app:A_extrapolation_ULIRG}

\begin{figure}[h!]
\centering
    \includegraphics[width=0.475\linewidth]{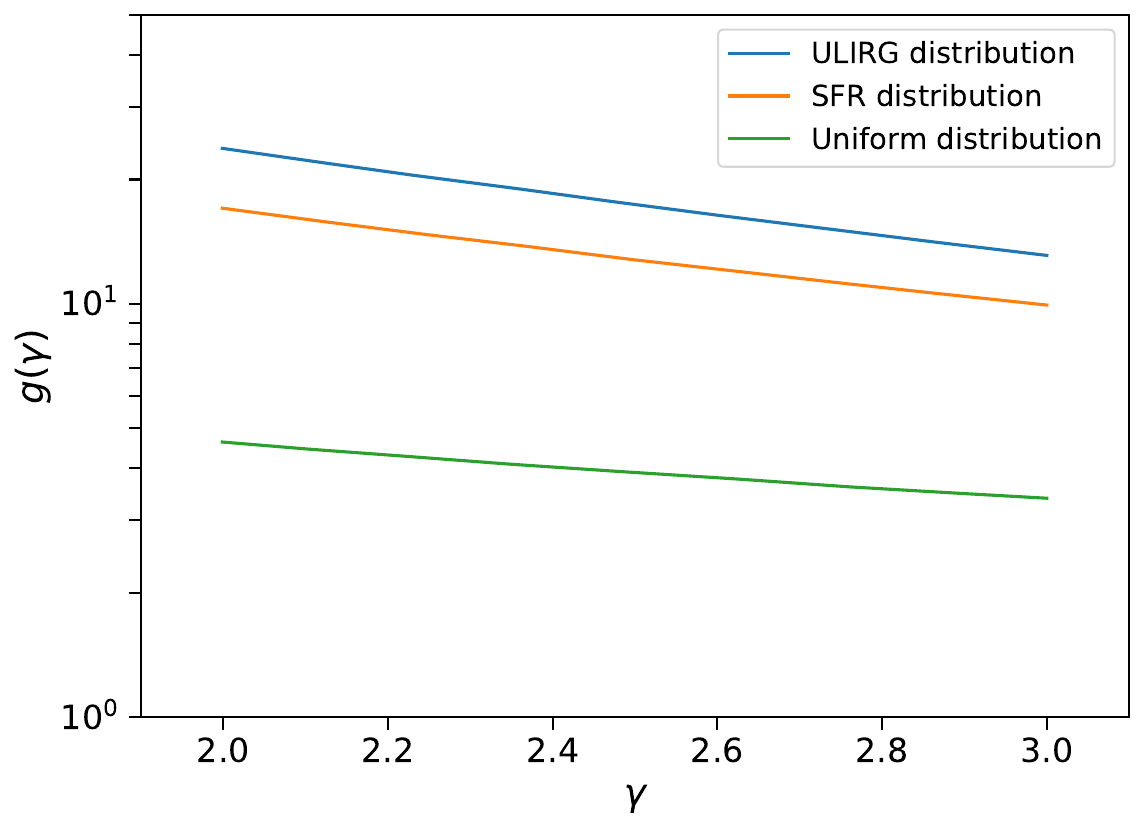}
    \includegraphics[width=0.5\linewidth]{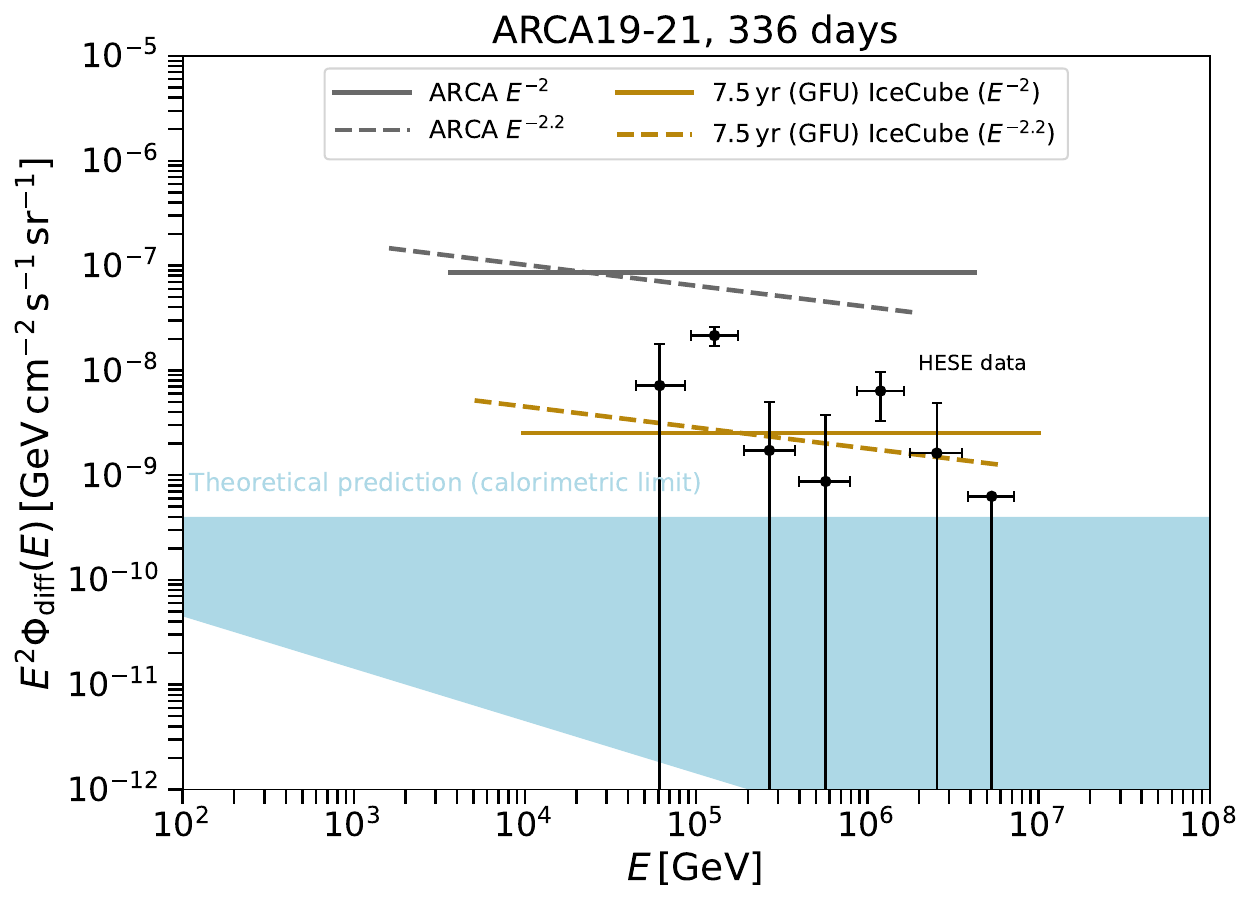}
    \caption{\textbf{Left}: $g(\gamma)$ as a function of the spectral index $\gamma$ for the three different redshift source distributions considered. \textbf{Right}: Extrapolated $90\%$ C.L. upper limits to the entire population of ULIRG assuming their redshift distribution. The IceCube upper limit~\cite{IceCube:2021waz} is shown, as well as the 7.5 year HESE flux~\cite{IceCube:2020wum} and the theoretical prediction. }
        \label{Fig:f_extrapolation}
\end{figure}

The analysed ULIRG catalogue represents the population of infrared galaxies up to redshift $z\le 0.13$. Therefore, the upper limit $\phi_s(E)$  can be used to constrain the maximal ULIRG flux with the relation
\begin{equation}\label{eq:diff_limit}
    \Phi_{\rm diff}(E,z=0.13) \le  \frac{\xi_s \phi_s (E)}{4\pi}\, ,
\end{equation}
where $\xi_s = 1.1$ takes into account the incompleteness of the catalogue, namely the effect of the limited sky coverage~\cite{IceCube:2021waz}. The factor $4\pi$ allows for estimating the upper limit on the flux per steradian. The limit of Eq.~\ref{eq:diff_limit} can be extrapolated to the entire ULIRG population in the Universe with 
\begin{equation}\label{eq:diff_limit_final}
    \Phi_{\rm diff}(E,z=4) = \Phi_{\rm diff} (E,z\le0.13) \frac{ \chi(\gamma,z \le 4)}{ \chi(\gamma, z \le 0.13)} \le \frac{\xi_s \phi_s (E)}{4\pi} \frac{ \chi(\gamma,z \le 4)}{ \chi(\gamma, z \le 0.13)}\, ,
\end{equation}
where 

\begin{equation}\label{eq:extrapolation_whole_sky}
    \chi(\gamma,z) =  \int_{0}^{z} dz^{'} \frac{f(z^{'})}{\sqrt{\Omega_{\Lambda} + (1+z^{'})^3 \Omega_{M}}} \cdot (1+z^{'})^{-\gamma}\, ,
\end{equation}
and is referred to as the cosmological factor which accounts  for the energy redshifting and the redshift distribution of sources~$f(z)$~\cite{IceCube:2021waz,Ahlers:2014ioa,Groth:2025aan}.   $\Omega_{M} = 0.31$ and $\Omega_{\Lambda} = 0.69$ are $\Lambda$CDM model parameters~\cite{Planck:2018vyg}.  The ULIRG distribution is  considered to be $f(z) = (1+z)^{4}$ for $z\le 1$ and constant for $1<z\le 4$ following Ref.~\cite{Vereecken:2020sqy}. For comparison, two other source distributions are considered: the uniform distribution $f(z) = 1$ and the Star Formation Rate (SFR) distribution $f(z) = (1+z)^{3.4}$ up to $z=1$ and $f(z) = (1+z)^{-0.3}$ for $1<z \le 4$~\cite{IceCube:2021waz,Hopkins:2006bw}.
In the left panel of Fig.~\ref{Fig:f_extrapolation} the factor $g(\gamma) = \frac{ \chi(\gamma,z \le 4)}{ \chi(\gamma, z \le 0.13)} $ is shown in terms of the spectral index for the three redshift distributions. The corresponding extrapolated upper limits for spectral index equal to 2.0, 2.2 and 2.4 are reported in Tab.~\ref{tab:extrapolation_benchmark}. \\
\newpage
\begin{table}[h!]
    \centering
    \begin{tabular}{|c|c|c|}
                \hline
                Source distribution & $\gamma$ & $E_0^2 \Phi_{\rm diff}^{90\%}(E_0) [\rm GeV\, \rm cm^{-2}\, \rm s^{-1}\, \rm sr^{-1}]$ \\
                \hline
                  & 2.0   & $1.7\times 10^{-7}$ \\
                ULIRG    & 2.2 & $2.4\times 10^{-7} $\\
                  & 2.4 & $3.9\times 10^{-7}$ \\
                \hline
                  & 2.0   & $1.2\times 10^{-7}$ \\
                SFR    & 2.2 & $1.8 \times 10^{-7} $\\
                  & 2.4 & $2.8\times 10^{-7}$ \\
         \hline
                  & 2.0   & $3.3\times 10^{-8}$ \\
                Uniform    & 2.2 & $4.8 \times 10^{-8} $\\
                  & 2.4 & $1.1 \times 10^{-7}$ \\
                  \hline
            \end{tabular}
    \caption{$90\%$ C.L. upper limits obtained for the diffuse flux of ULIRG considering three benchmark cases for the redshift distribution and the spectral index.}
    \label{tab:extrapolation_benchmark}
\end{table}

Finally, the extrapolated upper limits for the ULIRG distribution are shown in the right panel of Fig.~\ref{Fig:f_extrapolation}, assuming $E^{-2.0}$ and $E^{-2.2}$ spectra, together with the IceCube results~\cite{IceCube:2021waz} and the theoretical predictions, as a function of energy. The latter are obtained by extrapolating the predictions shown in Fig.~\ref{fig:results_1} using Eqs.~\ref{eq:diff_limit} and \ref{eq:diff_limit_final}, following the same procedure adopted for the reported upper limits.

\section{HBL stacking analysis: modelling the neutrino emission}
\label{app:B,HBLs}
\begin{figure}[h!]
    \centering
    \includegraphics[width=0.575\linewidth]{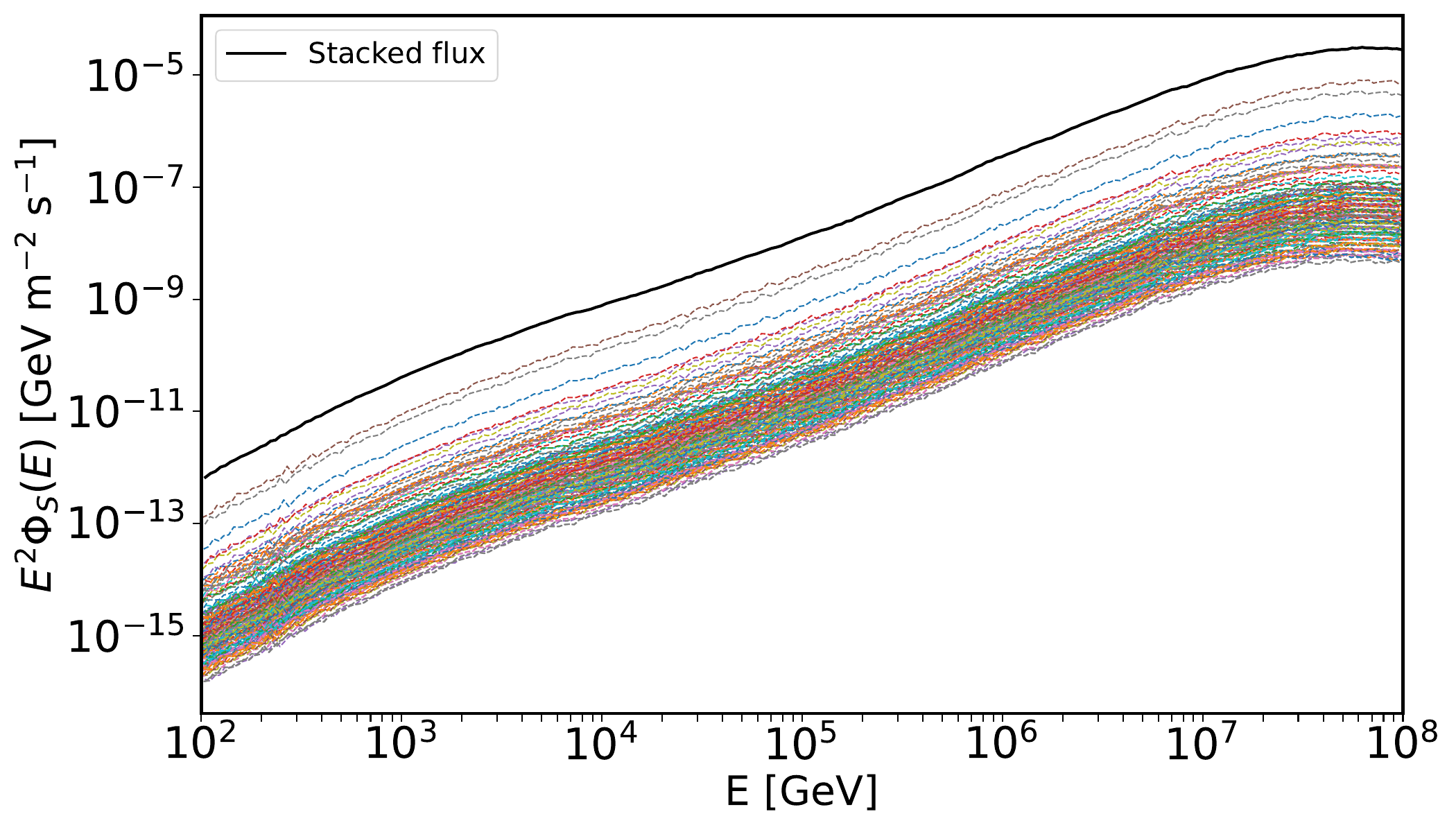}
    \caption{Expected neutrino fluxes for the 232 selected HBL sources (different colours refer to different sources in order to improve plot readability) and their stacked flux (black line). Neutrinos and antineutrinos are considered separately and their oscillated flavour composition at Earth is accounted for in the calculation.}
    \label{fig:template}
\end{figure}

In one-zone lepto-hadronic models, a magnetised compact region within the relativistic jet contains co-accelerated populations of relativistic electrons and protons. Neutrinos are produced through the decay of pions generated in proton-photon interactions. The high-energy component of the SED can arise, in addition to synchrotron-self-Compton from primary electrons (the leptonic contribution), also from synchrotron-pair cascades of secondary particles, neutral pion decay or proton synchrotron radiation (the hadronic contribution). The low-energy SED component is attributed to synchrotron emission from primary electrons. In this study, as a first step, the neutrino spectrum of the blazar PKS~2155-304 is computed using the LeHa-Paris code. PKS~2155-304 is a HBL object located in the Southern Hemisphere ($\delta = -30.22^{\circ}$) at redshift $z = 0.117$. It is selected as a representative HBL candidate, as it is the most luminous HBL in the Southern Sky, observed in gamma rays by H.E.S.S. in 2006~\cite{Aharonian:2007ig}. Its SED is studied by exploring a broad parameter space and comparing the results with multi-wavelength observational data from Ref.~\cite{Madejski:2016evb}, including joint NuSTAR and XMM-Newton observations that revealed a hard X-ray tail not readily explained by purely leptonic emission scenarios, further motivating lepto-hadronic modelling of this source. The optimisation of the modelling is detailed in~\cite{Carenini:2025Gv}, with the choice of the following parameters yielding the best-fit neutrino flux: an electron normalisation $K_e$, at a Lorentz factor \( \gamma_e = 1 \), of \(1.3 \times 10^4 \ \text{cm}^{-3} \); a maximum proton energy of \( \log_{10}(\gamma_p^{\text{max}}) = 7.4 \); a proton-to-electron ratio at \( \gamma_e = \gamma_p = 1 \) of \( \eta = 0.003 \); a spectral index for proton acceleration \( (\alpha_p) \) corresponding to 1.8; a magnetic field \( B = 0.036 \ \text{G} \); an emission region of radius \( R \simeq 7.9 \times 10^{16} \ \text{cm} \) and a break in the electron stationary energy distribution at \( \gamma_{\text{break}} = 6.3 \times 10^4 \). Additionally, a Doppler factor \( \mathcal{D}= 33 \) and a viewing angle of \( 0.1^\circ \) have been assumed~\cite{HESS:2019fhy}, corresponding to a bulk Lorentz factor \( \Gamma \simeq 16 \). 
\\
Rather than re-optimising the model parameters for each individual source, the expected neutrino flux for each blazar is obtained by rescaling the neutrino best-fit flux of PKS 2155-304 according to source-specific observables reported in the catalogue, thereby making the model predictive for a larger sample. Specifically, the synchrotron peak luminosity \( L_{\gamma,\text{peak}} \) serves as a proxy for the neutrino flux. This is motivated by the fact that synchrotron emission arises from the electron population, which is co-accelerated with protons under the same physical conditions. A more luminous synchrotron peak implies a denser photon field, and consequently, a potentially higher neutrino production efficiency. Then, the expected neutrino flux ($F_{\nu}^j$) for a generic source $j$ is estimated from
\begin{equation}\label{eq:neutrino_flux_rescaling}
F_{\nu}^j
=
F_{\nu}^{\mathrm{PKS}}
\,\frac{L_{\gamma,\mathrm{peak}}^j}{L_{\gamma,\mathrm{peak}}^{\mathrm{PKS}}}
\, ,
\end{equation}
where $F_{\nu}^{\rm PKS}$ is the PKS 2155-304 neutrino flux, $\frac{L_{\gamma,\rm peak}^j}{L_{\gamma,\rm peak}^{\rm PKS}}$ is the ratio between the synchrotron peak luminosity of the source $j$ with the one of PKS~2155-304. Another argument supporting this scaling is the assumption that all sources share the same Doppler factor. While this is clearly a simplification, it enables a consistent comparison across the sample. Because both photon and neutrino luminosities scale identically with this parameter, the ratio of neutrino luminosities between sources can be approximated by the ratio of their synchrotron peak fluxes. The final neutrino flux calculations take into account the different distances of the sources and also the energy redshifting as reported in Ref.~\cite{Carenini:2025Gv}. \\
Thus, neutrino fluxes for the 232 selected HBL sources have been derived: their stacked contribution is depicted in Fig.~\ref{fig:template} together with the predicted fluxes of individual blazars. The fluxes are displayed in the energy range from $100~\rm GeV$ to $100~ \rm PeV$, relevant for high-energy neutrino studies with the KM3NeT/ARCA detector. The calculation accounts for neutrinos and antineutrinos, as well as their mixed flavour composition at Earth due to neutrino oscillations over cosmic distances. Because of the adopted rescaling procedure, all the templates share a common neutrino spectral shape, featuring a pronounced peak at the level of $10^{17}$ eV. This neutrino spectral shape is then shifted in energy due to the redshift.

\vspace{-0.5 cm}

\section{EHBL stacking analysis: modelling the neutrino emission}
\label{app:C,EHBLs}
\begin{figure}[H]
    \centering
    \includegraphics[width=0.57\linewidth]{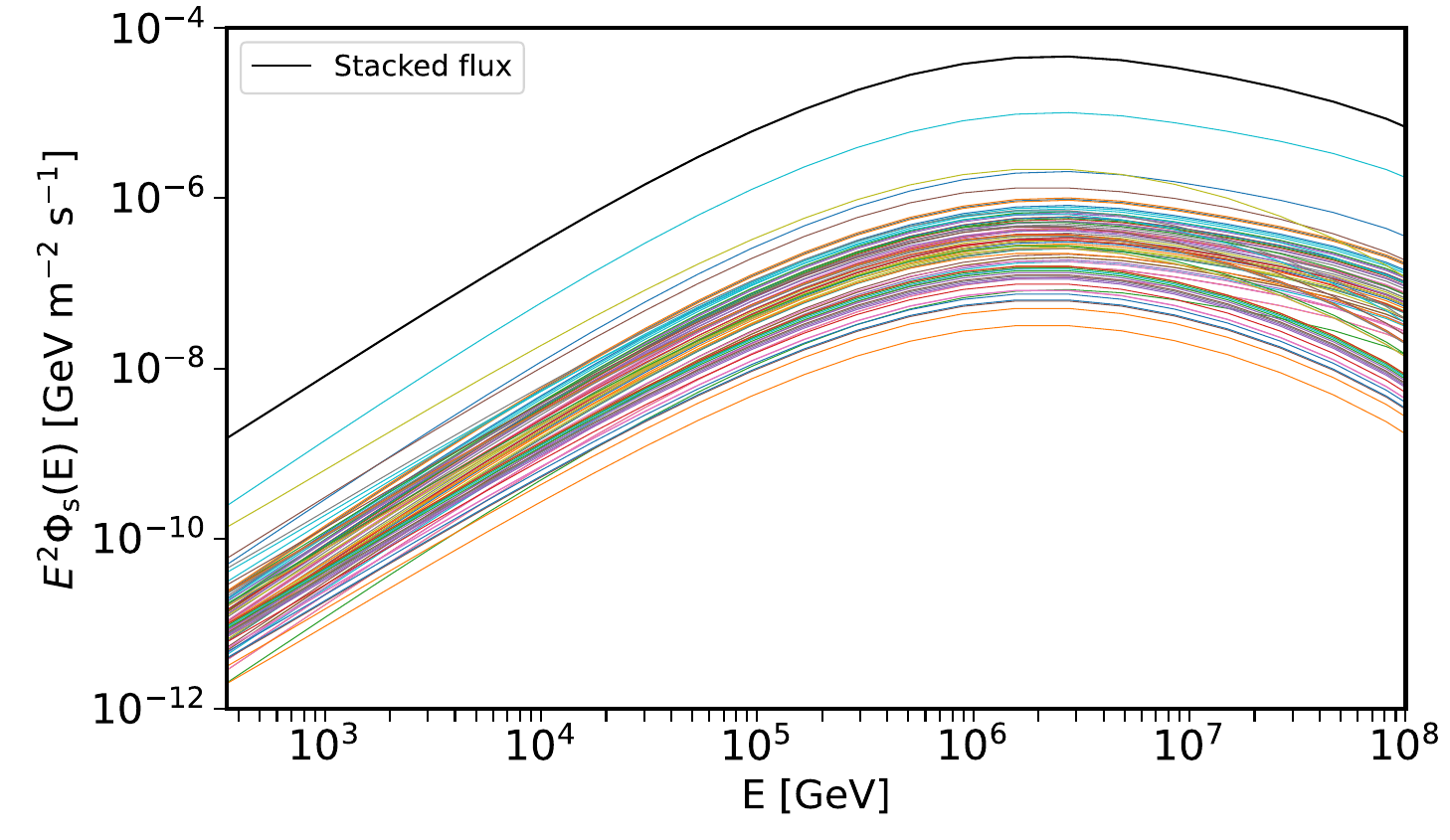}
    \caption{ Expected neutrino fluxes for the 88 selected EHBL sources (different colours refer to different sources in order to improve plot readability) and their stacked flux (black line). Neutrinos and antineutrinos are considered separately and their oscillated flavour composition at Earth is accounted for in the calculation.}
    \label{fig:EHSP}
\end{figure}

The one-zone scenario is also applied to the EHBL sample, with relativistic particles (primary electrons and protons) injected into the region at a constant rate with broken power-law distributions. The neutrino emission is predicted adapting the open-source lepto-hadronic code LeHaMoC~\cite{Lehamoc}, taking into account all the relevant physical processes for protons, such as the pair and photopion production mechanisms. 
As outlined in the main text, the average SED of extreme blazars is dominated by leptonic processes by primary electrons~\cite{Keivani:2018rnh}, making hadronic processes totally subdominant for gamma-ray production. Therefore, the SEDs are fitted considering a pure leptonic model and the hadronic contribution is inserted afterwards, considering the same parameters for protons and electrons, fixing the  baryonic loading factor to $10^4$. The interest in EHBL is related to their higher peak frequencies for the first bump of their SED compared to HBL. In fact, the photons emitted as synchrotron radiation are the target for the photopion process, mainly producing neutrinos in energy range of $100 \, \rm TeV - 10\, \rm PeV$, where KM3NeT/ARCA is more sensitive. \\
As for the HBL case, instead of optimising the model parameters for each individual source, three sources have been chosen as prototypes: 3HSP~J034923.2-11, 3HSP~J023248.6+20 (known as 1ES 0229+200, the most extreme source in the catalogue) and 3HSP~J235907.9-30. The three sources are representative of the variability of the measured synchrotron peak frequencies and define three distinct ranges: 
$ \nu_{\text{peak}}^{\text{syn}} \leq 10^{17.1}\,\text{Hz}$, 
$10^{17.1}\,\text{Hz} < \nu_{\text{peak}}^{\text{syn}} \leq 10^{17.9}\,\text{Hz}$, and 
$ \nu_{\text{peak}}^{\text{syn}} > 10^{17.9}\,\text{Hz}$. 
For the scaling, the integrated flux~$(F_{\gamma})$ in the $50\ \rm MeV -1\ \rm TeV$ band is chosen, representing the sensitivity range of the Fermi-LAT telescope. In order to calculate the neutrino flux for each source $j$, a weight is evaluated, which takes into account the ratio between the integrated flux~$(F_{\gamma})$ in the $50\ \rm MeV -1\ \rm TeV$ and  the benchmark of the prototype source ($F_\gamma ^*$)~\cite{Fermi-LAT:2019yla}, as reported in the following equation:

\begin{equation}
F_{j,\nu}= F_{\nu}^{*}\bigg(\frac{F_{j,\gamma}}{F_\gamma^*}\bigg)\, ,
\label{eq:flux}
\end{equation}
where $F_{\nu}^{*}$ represents the prototype neutrino flux. As for the HBL case, the final neutrino flux calculation also accounts for the energy redshifting. This approach allows for a consistent and homogeneous scaling factor across all sources, ensuring the application of a uniform criterion. Indeed, X-ray data are not always available and are often obtained with different instruments. Consequently, using the X-ray integrated flux from the 3HSP catalogue to scale the expected neutrino flux would not be reliable.\\
To account for the effects introduced by the assumptions underlying the scaling procedure, an uncertainty is assigned to the stacked neutrino flux. It is estimated by comparing the expected neutrino flux for 3HSP~J034923.2-11, obtained from the direct fit, with the fluxes reconstructed by applying the rescaling method using the other two prototype EHBL as reference sources. The same validation procedure was performed for all three prototype blazars. 
In Fig.~\ref{fig:EHBLuL} the obtained stacked flux is reported together with its uncertainty band. The neutrino templates for the 88 selected EHBL sources and their stacked (summed) contribution are shown in Fig.~\ref{fig:EHSP}. In analogy to the case of HBL, the calculation accounts for neutrinos and antineutrinos, as well as their mixed flavour composition at Earth due to neutrino oscillations over cosmic distances.

\newpage

\bibliography{bib}{}
\bibliographystyle{unsrt}

\end{document}